\documentclass[journal=jacsat,manuscript=article]{achemso}
\SectionNumbersOn
\usepackage[version=3]{mhchem} % Formula subscripts using \ce{}
\usepackage{amsmath}% 
\usepackage{graphicx}%
\usepackage{xcolor}%
\usepackage{dcolumn}%
\usepackage{bm}%
\usepackage{ulem}
\usepackage[uline]{hhtensor}
\usepackage{setspace}
\usepackage{siunitx}
\usepackage{lineno,hyperref}
\usepackage{subcaption}

\newcommand{\bra}{\langle}
\newcommand{\ket}{\rangle}

\author{Marco Marchetta \footnote{M.M. and C.M. contributed equally to this paper.} }
\affiliation{Dipartimento di Scienze Chimiche e Farmaceutiche, Universit\`a di Trieste, Trieste 34127, Italy}
\author{Chiara Morassut}
\affiliation{Laboratoire de Chimie Th\'eorique, Sorbonne Universit\'e, CNRS, Paris, F-75005, France}
\alsoaffiliation{Dipartimento di Scienze Chimiche e Farmaceutiche, Universit\`a di Trieste, Trieste 34127, Italy}
\author{Julien Toulouse}
\affiliation{Laboratoire de Chimie Th\'eorique, Sorbonne Universit\'e, CNRS, Paris, F-75005, France}
\alsoaffiliation{Institut Universitaire de France, F-75005 Paris, France}
\author{Emanuele Coccia}
\email{ecoccia@units.it}
\affiliation{Dipartimento di Scienze Chimiche e Farmaceutiche, Universit\`a di Trieste, Trieste 34127, Italy}
\author{Eleonora Luppi}
\email{eleonora.luppi@sorbonne-universite.fr}
\affiliation{Laboratoire de Chimie Th\'eorique, Sorbonne Universit\'e, CNRS, Paris, F-75005, France}

\title[Orbital decomposition in HHG]{Time-dependent ab initio molecular-orbital decomposition for high-harmonic generation spectroscopy}

\begin{document}
\begin{abstract}
We propose a real-time time-dependent ab initio approach within a configuration-interaction-singles ansatz
to decompose the high-harmonic generation (HHG) signal of molecules in terms of individual molecular-orbital (MO) contributions.
Calculations have been performed by propagating the time-dependent Schr\"odinger equation with complex energies, in order to account for ionization of the system, and by using tailored Gaussian basis sets for high-energy and continuum states. We have studied the strong-field electron dynamics and the HHG spectra in aligned CO$_2$ and H$_2$O molecules. Contribution from MOs in the strong-field dynamics depends on the interplay between the MO ionization energy and the coupling between the MO and the laser-pulse symmetries. Such contributions characterize different portions of the HHG spectrum, indicating that the orbital decomposition encodes nontrivial information on the modulation of the strong-field dynamics. Our results correctly reproduce the MO contributions to HHG for CO$_2$ as described in literature's experimental and theoretical data, and lead to an original analysis of the role of the highest occupied molecular orbitals HOMO, HOMO-1, and HOMO-2 of H$_2$O according to the polarization direction of the laser pulse.
 \end{abstract}

\section{Introduction}

High-harmonic generation (HHG) is a highly nonlinear phenomenon resulting from the interaction of an intense infrared laser pulse with atom, molecules \cite{Marangos:2016:13/132001}, solids \cite{li20a,par22} and, recently, also liquids \cite{NC-9-3723,neu22,dichiara2009investigation}. The HHG process can be understood by the three-step model (3SM) where three important assumptions are made: (i) an electron escapes from the field of the nuclei through tunnel ionization associated with the strong laser field, (ii) it is accelerated in the electronic continuum until the sign of the laser field changes, (iii) whereupon the electron is reaccelerated back to the nuclei where it recombines by emission of a photon which energy is a harmonic of the laser-field frequency. \cite{cork93prl,lewe+pra94,san06}.

HHG spectra are widely employed to investigate the electronic structure and dynamics in atomic and molecular systems \cite{Peng2019,science.1163077,FD-194-369,PRL-114-153901,Angew-57-5228-2018,voz16,nis17,PRL-117-093902-2016,PRL-95-153902}. In fact, the physical mechanisms of ionization and recombination occurring during the strong-field dynamics involve the interaction between different molecular orbitals (MOs) and therefore the HHG spectrum contains information about the structures and the electron dynamics of the molecules \cite{Smirnova2009}. The structural and dynamic information can be extracted from molecules by high-harmonic interferometry, from which the role of different MOs can be revealed, together with multielectron contributions to HHG. \cite{Smirnova2009,Uzan2020} Tomographic imaging of MOs has been possible from HHG \cite{itat04nat,Haessler:2010hb,DIVEKI2013121,Peng2019,FD-171-133,Nat-Phys-7-822-2011}, stimulating the study of attosecond charge migration in molecules. \cite{He2022}

For a long time, it has often been assumed that the features of the HHG spectra reflected only the spatial distribution of the highest occupied molecular orbital (HOMO). Now, it is instead well established that the ionization and recombination processes in HHG can involve other deeper MOs than the HOMO. Indeed, other MOs than the HOMO can have a spatial symmetry that is more favorable for the ionization and recombination, when coupled to a given laser-pulse polarization. Various experiments and theoretical calculations pointed out multiple-MO contributions to HHG \cite{wor10,Smirnova2009,tut20,Marangos:2016:13/132001,lup23,mor24,PhysRevA.93.013422,mcfarland2008high,le2009uncovering,Li:13}. 

An important example of this is provided by the linear molecule CO$_{2}$. The HHG spectrum of CO$_{2}$ has a minimum that has been deeply studied over the years. \cite{wor10,Smirnova2009,PhysRevA.83.053409,PRL-114-153901,rub18,pau21,jin20,jina20,tut22,shu22} It is now well understood that this minimum is due to multiple-MO contributions to the HHG emission, i.e. it has a dynamical nature \cite{rub18,shu22,tut22}. 
HHG from gas-phase H$_2$O has also been explored, as a prototype of a nonlinear molecule.~\cite{wong2010high,zha11,ren2023molecular} 
Despite the central role of water in practically any field of chemistry and biology, only recently attention has been devoted to the HHG spectroscopy of aligned gas-phase water molecules. \cite{ren2023molecular}

A clear understanding of the mechanisms that control the strong-field electron dynamics in HHG is still challenging. Developing theoretical methods to analyze the dynamics of electrons during the ionization and the recombination process can give a strong impulse in understanding HHG in molecular systems.\cite{ec21,woz22,PhysRevA.93.013422}. 
Recently, using real-time time-dependent configuration-interaction singles (RT-TD-CIS)~\cite{Rohringer:2006kp,Krause:2007dm,PRA-82-023406,lupp+12mol,white15,app8030433,bed19,ec21,woz22}, some of us studied the role of MOs in the HHG of aligned gas-phase uracil \cite{lup23} and randomly aligned bromoform molecules \cite{mor24}. For both molecules, it was found that MOs other than the HOMO are fundamentally important in explaining the features of the HHG signals.

The approach proposed in Refs. \citenum{lup23,mor24} provided an approximated decomposition of the time-dependent dipole moment as a sum of the time-dependent dipole moments calculated from each occupied orbital, weighted by the contributions from the ground and excited states. Then, by Fourier transforming the contribution of the $i$-th MO, we were able to determine which MO(s) mainly contributed to the HHG signal.

In the present work, we improve our methodology by proposing an exact decomposition of the time-dependent dipole moment into MO contributions, where for each orbital, not only the ground-excited state contributions but also the depletion of the ground state and the excited-excited state contributions are included. This exact approach can, therefore, distinguish the role of each MO, identify the interferences between them, and extract the ground/excited-state contributions during the strong-field dynamics in the HHG process.

We have applied our method to calculate the HHG spectra of aligned CO$_{2}$ and of H$_2$O with different linearly polarized laser pulses, analyzing the contribution of each MO to the HHG spectrum.

The paper is organized as follows: our theoretical approach is derived in Section \ref{theo}, computational details are given in Section \ref{det}, results for CO$_{2}$ and H$_2$O are shown and discussed in Section \ref{res}, conclusions and perspectives for future work are in Section \ref{con}.
  
\section{Theory}
\label{theo}

\subsection{Real-time propagation}

The time-dependent Schr\"odinger equation (TDSE) for a spectroscopic target irradiated by an external time-dependent electric field is 
\begin{equation}
i\frac{ \partial \vert \Psi(t) \rangle}{\partial t} = \left(\hat{H}_0 + \hat{V}(t) \right)\vert\Psi(t) \rangle,
\label{tdse}
\end{equation}
where $\hat{H}_0$ is the time-independent field-free Hamiltonian and $\hat{V}(t)$ is the time-dependent potential operator.  \cite{Bandrauk:2009ig,PRA-81-023411-2010,PRA-81-063430-2010,PRA-84-035402-2011,coccia16b}
The time-independent field-free Hamiltonian is:
\begin{equation}
\hat{H}_0=\hat{T} + \hat{W}  + \hat{V}_\text{ne},
\label{nofield}
\end{equation}
where $\hat{T}$ is the electron kinetic-energy operator, $\hat{W}$ is the electron-electron Coulomb repulsion operator, and $\hat{V}_\text{ne}$ is the static nuclei-electron Coulomb attraction operator. 

The interaction between the molecule and the time-dependent external laser field in the semi-classical dipole approximation in the length gauge is
\begin{equation}
\hat{V}(t) = -\hat{\boldsymbol{\mu}}\cdot{\bf E}(t),
\end{equation}
where $\hat{\boldsymbol{\mu}}$ is the molecular dipole operator and ${\bf E}(t)$ is the electric field of the laser pulse.
We consider a linearly-polarized electric field ${\bf E}(t)$ along the $\alpha$ axis ($\alpha=x$, $y$ or $z$), representing a laser pulse, 
\begin{equation}
{\bf E}(t) = {\cal E}_0 {\bf n}_\alpha \sin(\omega_{0} t + \phi) f(t),
\label{e-field}
\end{equation}
where ${\cal E}_0$ is the maximum field strength, ${\bf n}_\alpha$ is a unit vector along the $\alpha$ axis, $\omega_{0}$ is the carrier frequency, $\phi$ is a phase, and $f(t)$ is the envelope function chosen as
\begin{equation} 
f(t) = 
\begin{cases}  
\cos^{2}( \frac{\pi}{2\sigma} (\sigma - t) )   & \text{if }  \vert t - \sigma \vert \le \sigma\\ 
0 & \text{otherwise},
\end{cases} 
\label{coslaser}
\end{equation}
where $\sigma$ is the width of the field envelope.

To solve Eq.~(\ref{tdse}) in a finite basis set, the wave function $\vert\Psi(t)\rangle$ is expanded in a discrete basis of eigenstates $\{\Psi_n\}_{n\geq 0}$ of the field-free Hamiltonian $\hat{H}_0$ composed of the ground state ($n=0$) and excited states ($n>0$)
\begin{equation}
\vert\Psi(t)\rangle = c_{0}(t) \vert \Psi_{0} \rangle + \sum_{n>0} c_{n}(t) \vert \Psi_{n} \rangle ,
\label{wft}
\end{equation} 
where ${c}_{n}(t)$ are complex-valued time-dependent coefficients. Inserting Eq.~(\ref{wft}) into Eq.~(\ref{tdse}), and projecting on the eigenstates $\bra\Psi_{m}\vert$ with $m\ge0$, gives the evolution equations of the coefficients which in matrix form are
\begin{equation}
i\frac{\partial {\bf c}(t)}{\partial t} = \left( {\bf H}_0 + {\bf V}(t) \right) {\bf c}(t),
\label{CIequation}
\end{equation}
where ${\bf c}(t)$ is the column vector of the coefficients ${c}_{n}(t)$, ${\bf H}_0$ is the field-free Hamiltonian diagonal matrix with elements ${\bf H}_{0,mn} = \bra \Psi_m \vert \hat{H}_0 \vert \Psi_n \ket = E_n \delta_{mn}$ (where $E_n$ is the energy of the eigenstate $n\ge0$), and ${\bf V}(t)$ is the laser-interaction non-diagonal matrix with elements ${\bf V}_{mn}(t)= \bra \Psi_m\vert\hat{V}(t) \vert \Psi_n\ket$. The initial wave function 
at $t=t_\text{i}=0$ is chosen to be the field-free ground state, i.e. $c_n(t_\text{i}) = \delta_{n0}$. To solve Eq.~(\ref{CIequation}), time is discretized and the split-propagator approximation is used, \cite{Castro:2004hk} which reads as
\begin{equation}
{\bf c}(t + \Delta t) =  e^{-i {\bf V}(t) \frac{\Delta t}{2}}  e^{-i {\bf H}_0 \Delta t} e^{-i {\bf V}(t) \frac{\Delta t}{2}}  {\bf c}(t),
\label{coefficients}
\end{equation}
where $\Delta t$ is the time step of the propagation. Since the matrix ${\bf H}_0$ is diagonal, $e^{-i {\bf H}_0 \Delta t}$ is also a diagonal matrix with elements $e^{-i E_n \Delta t}\delta_{mn}$. The exponential of the non-diagonal matrix ${\bf V}(t)$ is calculated as 
\begin{equation}
e^{-i {\bf V}(t) \Delta t} = {\bf U}^{\dagger} \; e^{-i {\bf V}_\text{d} (t) \Delta t} \; {\bf U},
\end{equation}
where ${\bf U}$ is the (time-independent) unitary matrix describing the change of basis between the original eigenstates of $\hat{H}_0$ and a basis in which $\hat{V}(t)$ is diagonal, i.e. ${\bf V}_\text{d}(t) = {\bf U} {\bf V}(t){\bf U}^{\dagger}$. \cite{lupp+12mol,coccia16b}

Once the time-dependent wave-function $\vert\Psi(t)\ket$ is known, the time-dependent dipole moment ${\boldsymbol \mu}(t) = \langle \Psi(t) \vert \hat{\boldsymbol  \mu} \vert  \Psi(t)  \rangle$ is computed from which, by taking the Fourier transform, the HHG power spectrum is obtained
\begin{equation}
P(\omega) = \bigg\vert \frac{1}{t_\text{f} - t_\text{i}}\int^{t_\text{f}}_{t_\text{i}} W(t) {\boldsymbol  \mu}(t) \cdot {\bf n}_{\alpha'}  \; e^{-i\omega t} dt \bigg\vert^{2},
\label{eq:spectrum}
\end{equation}  
where $t_\text{i}$ and $t_\text{f}$ are the initial and final propagation times, $W(t)$ a window function, and ${\bf n}_{\alpha'}$ is the unit vector in the direction of emission of the HHG signal, which can be in general different from the polarization unit vector ${\bf n}_{\alpha}$.

\subsection{TD-CIS wave-function ansatz}

In this work, we use a TD-CIS ansatz for the time-dependent wave function, i.e. the ground-state is approximated by a single reference Slater determinant 
\begin{equation}
\vert \Psi_0 \rangle \approx  \vert \Phi_0 \rangle, 
\label{Phi0}
\end{equation} 
and the excited states by linear combinations of singly excited Slater determinants
\begin{equation}
\vert \Psi_{n} \rangle  \approx  \sum_{i}^\text{occ} \sum_{a}^\text{vir} r_{i,n}^{a} \vert \Phi_{i}^{a} \rangle, \;\; \text{for}\; n>0,
\label{wfcis}
\end{equation} 
where $i$ and $a$ run over (real-valued) occupied and virtual (spin) orbitals, respectively,  $\vert \Phi_{i}^{a} \rangle$ is a singly excited Slater determinant with respect to $\vert \Phi_0 \rangle$, and $r_{i,n}^{a}$ are real-valued coefficients.

The TD-CIS wave-function ansatz is quite general, since different reference ground-state Slater determinant $\vert \Psi_0 \rangle$ and singly excited states $\vert \Psi_{n} \rangle$ can be used. We use two possibilities: (1) the Hartree-Fock (HF) ground-state determinant and CIS excited states, and (2) the Kohn-Sham (KS) ground-state determinant and linear-response TDDFT excited states in the Tamm-Dancoff approximation (TDA).

\subsubsection{HF ground state and CIS excited states}
The first possibility is to use the HF ground-state determinant of the field-free system for $\vert \Phi_0 \rangle$ and the corresponding CIS excited states for $\vert \Psi_{n} \rangle$, resulting in a time-dependent method named RT-TD-CIS~\cite{li05,Schlegel:2007ke,Sonk:2011ch,kra15,kra15b,hoe15,hoe17,hoe17b,hoe18,bed19,SAALFRANK202015,hoe21,pau21}. The CIS excited states are found by solving the CIS eigenvalue matrix equation~\cite{Dreuw:2005gz}
\begin{equation}
{\bf A}^\text{CIS} \; {\bf X}_n = \omega^{\text{CIS}}_n \;{\bf X}_n,
\label{matrixCIS}
\end{equation}
where the matrix elements of ${\bf A}^\text{CIS}$ are given by
\begin{equation}
A_{ia,jb}^\text{CIS} = (\epsilon^{\text{HF}}_a - \epsilon^{\text{HF}}_i) \delta_{ij} \delta_{ab} + \bra aj \vert \hat{w}_\text{ee}\vert ib \ket  -  \bra aj \vert \hat{w}_\text{ee}\vert bi \ket ,
\label{rtcis}
\end{equation}
where $i,j$ and $a,b$ run over occupied and virtual HF orbitals, respectively, $\epsilon^{\text{HF}}_i$ and  $\epsilon^{\text{HF}}_a$ are HF occupied and virtual orbital energies, respectively, and $\bra aj \vert \hat{w}_\text{ee}\vert ib \ket$  and $\bra aj \vert \hat{w}_\text{ee}\vert bi \ket$ are the Coulomb two-electron integrals associated with the Hartree and Fock exchange terms, respectively. Solving Eq.~(\ref{matrixCIS}) gives the column eigenvectors ${\bf X}_n$ containing the CIS excited-state coefficients $X_{n,ia}=r_{i,n}^{a}$, and the CIS excitation energies $\omega^{\text{CIS}}_n$ leading to the excited-state energies $E_n = E^{\text{HF}}_0 + \omega^{\text{CIS}}_n$ where $E^{\text{HF}}_0$ is the ground-state HF energy.

\subsubsection{KS ground state and linear-response TDDFT-TDA excited states}

The second possibility is to use the KS ground-state determinant of the field-free system for $\vert \Phi_0 \rangle$ and the corresponding linear-response TDDFT-TDA excited states for $\vert \Psi_{n} \rangle$. In this case, the excited states are found by solving the linear-response TDDFT-TDA eigenvalue matrix equation, in the adiabatic approximation,
\begin{equation}
{\bf A}^\text{TDDFT-TDA}\; {\bf X}_n = \omega^{\text{TDDFT-TDA}}_n \; {\bf X}_n,
\label{tddft-tda}
\end{equation}
where the matrix elements of ${\bf A}^\text{TDDFT-TDA}$ are given by
\begin{eqnarray}
A_{ia,jb}^\text{TDDFT-TDA} = (\epsilon^{\text{KS}}_a - \epsilon^{\text{KS}}_i) \delta_{ij} \delta_{ab} +\! \bra aj \vert \hat{w}_\text{ee}\vert ib \ket  + \bra aj \vert \hat{f}_{\text{xc}} \vert ib \ket ,
\label{tddft}
\end{eqnarray}
where $\epsilon^{\text{KS}}_i$ and $\epsilon^{\text{KS}}_a$ are KS occupied and virtual orbital energies, respectively, and $\bra aj \vert \hat{w}_\text{ee}\vert ib \ket$ and $\bra aj \vert \hat{f}_{\text{xc}} \vert ib \ket $ are the two-electron integrals associated with the Hartree and the DFT exchange-correlation kernel, respectively. Solving Eq.~(\ref{tddft-tda}) gives the column eigenvectors ${\bf X}_n$ containing the TDDFT-TDA excited-state coefficients $X_{n,ia}=r_{i,n}^{a}$, and the  TDDFT-TDA excitation energies $\omega^{\text{TDDFT-TDA}}_n$ leading to the excited-state energies $E_n = E^{\text{DFT}}_0 + \omega^{\text{TDDFT-TDA}}_n$ where $E^{\text{DFT}}_0$ is the ground-state DFT energy.

In this work, we use the LC-$\omega$PBE exchange-correlation approximation \cite{wpbe2,wpbe1} where the kernel is
\begin{eqnarray}
f^{\text{LC-$\omega$PBE}}_{\text{xc}} = f^{\text{lrHF,$\omega$}}_{\text{x}} + f^{\text{srPBE,$\omega$}}_{\text{x}} + f^{\text{PBE}}_{\text{c}},
\end{eqnarray}
where $\omega$ is the range-separation parameter, $f^{\text{lrHF,$\omega$}}_{\text{x}}$ is the long-range HF exchange kernel, $f^{\text{srPBE,$\omega$}}_{\text{x}}$ is the short-range PBE exchange kernel, and $f^{\text{PBE}}_{\text{c}}$ is the PBE correlation kernel. The resulting real-time propagation method will be named RT-TD-CIS-LC-$\omega$PBE.

\subsection{MO decomposition of the dipole moment and of the HHG spectrum}

For interpretational purposes, it is interesting to decompose the  time-dependent dipole moment  ${\boldsymbol \mu}(t) = \langle \Psi(t) \vert \hat{\boldsymbol  \mu} \vert  \Psi(t)  \rangle$ and the corresponding HHG spectrum in Eq.~(\ref{eq:spectrum}) into single MO contributions.

Starting from the time-dependent wave function in Eq.~(\ref{wft}), we first decompose the time-dependent dipole moment in three contributions, 
\begin{eqnarray}
{\boldsymbol \mu}(t) = {\boldsymbol \mu}_\text{G}(t) + {\boldsymbol \mu}_\text{G-E}(t) + {\boldsymbol \mu}_\text{E-E}(t), 
\label{}
\end{eqnarray}
where ${\boldsymbol \mu}_\text{G}(t)$ is the ground-state contribution
\begin{eqnarray}
{\boldsymbol \mu}_\text{G}(t) = \vert c_{0}(t) \vert^2 \langle \Psi_{0} \vert \hat{{\boldsymbol \mu}} \vert \Psi_{0} \rangle,
\label{}
\end{eqnarray}
${\boldsymbol \mu}_\text{G-E}(t)$ is the ground-state/excited-state contribution
\begin{eqnarray}
{\boldsymbol \mu}_\text{G-E}(t) = \sum_{n>0} \Bigl( c_0^*(t) c_n(t) \langle \Psi_{0} \vert \hat{{\boldsymbol \mu}} \vert \Psi_{n} \rangle + c_n^*(t) c_0(t) \langle \Psi_{n} \vert \hat{{\boldsymbol \mu}} \vert \Psi_{0} \rangle  \Bigl),
\label{}
\end{eqnarray}
and ${\boldsymbol \mu}_\text{E-E}(t)$ is the excited-state/excited-state contribution
\begin{eqnarray}
{\boldsymbol \mu}_\text{E-E}(t) = \sum_{n>0} \sum_{m>0} c_m^*(t) c_n(t) \langle \Psi_{m} \vert \hat{{\boldsymbol \mu}} \vert \Psi_{n} \rangle.
\label{}
\end{eqnarray}
For a TD-CIS wave-function ansatz [Eqs.~(\ref{Phi0}) and~(\ref{wfcis})], the matrix elements can be expressed in terms of the dipole-moment integrals over the MOs
\begin{eqnarray}
\langle \Psi_{0} \vert \hat{{\boldsymbol \mu}} \vert \Psi_{0} \rangle = \sum_{i}^\text{occ}\langle i \vert \hat{{\boldsymbol \mu}} \vert i \rangle,
\label{}
\end{eqnarray}
\begin{eqnarray}
\langle \Psi_{0} \vert \hat{{\boldsymbol \mu}} \vert \Psi_{n} \rangle = \sum_{i}^\text{occ} \sum_{a}^\text{vir} r_{i,n}^a \langle i \vert \hat{{\boldsymbol \mu}} \vert a \rangle,
\label{}
\end{eqnarray}
\begin{eqnarray}
\langle \Psi_{m} \vert \hat{{\boldsymbol \mu}} \vert \Psi_{n} \rangle = \sum_{i}^\text{occ} \sum_{j}^\text{occ} \sum_{a}^\text{vir} \sum_{b}^\text{vir} r_{i,m}^a  r_{j,n}^b \left( \sum_{k}^\text{occ} \langle k \vert \hat{{\boldsymbol \mu}} \vert k \rangle \delta_{ij} \delta_{ab} + \langle a \vert \hat{{\boldsymbol \mu}} \vert b \rangle \delta_{ij} - \langle j \vert \hat{{\boldsymbol \mu}} \vert i \rangle \delta_{ab} \right).
\label{}
\end{eqnarray}
By factorizing the sum over the occupied orbitals $i$, we obtain a MO decomposition of the dipole moment~\cite{lup23,mor24}
\begin{equation}
\boldsymbol{\mu}(t)  =  \sum_i^\text{occ}\boldsymbol{\mu}_{i}(t), 
\end{equation}
with
\begin{equation}
\boldsymbol{\mu}_i(t)  =  \boldsymbol{\mu}_{\text{G},i}(t) + \boldsymbol{\mu}_{\text{G-E},i}(t) +  \boldsymbol{\mu}_{\text{E-E},i}(t),
\label{muGGEEE}
\end{equation}
where
\begin{equation}
\boldsymbol{\mu}_{\text{G},i}(t) = \vert c_0(t) \vert^2 \langle i \vert \hat{{\boldsymbol \mu}} \vert i \rangle,
\end{equation}
\begin{equation}
\boldsymbol{\mu}_{\text{G-E},i}(t) = \sum_{n>0} \sum_{a}^\text{vir}  \left( c_0^*(t) c_n(t) + c_n^*(t) c_0(t) \right)  r_{i,n}^a \langle i \vert \hat{{\boldsymbol \mu}} \vert a \rangle,
\label{ge}
\end{equation}
and
\begin{eqnarray}
\boldsymbol{\mu}_{\text{E-E},i}(t) &=& \sum_{n>0} \sum_{m>0} \sum_{k}^\text{occ} \sum_{a}^\text{vir} c_m^*(t) c_n(t)   r_{i,m}^a  r_{i,n}^a  \langle k \vert \hat{{\boldsymbol \mu}} \vert k \rangle 
\nonumber\\
&&+\sum_{n>0} \sum_{m>0} \sum_{a}^\text{vir} \sum_{b}^\text{vir} c_m^*(t) c_n(t)  r_{i,m}^a  r_{i,n}^b \langle a \vert \hat{{\boldsymbol \mu}} \vert b \rangle 
\nonumber\\
&&-\sum_{n>0} \sum_{m>0}\sum_{j}^\text{occ} \sum_{a}^\text{vir} c_m^*(t) c_n(t)   r_{i,m}^a  r_{j,n}^a  \langle j \vert \hat{{\boldsymbol \mu}} \vert i \rangle.
\end{eqnarray}

In Refs.~\cite{lup23, mor24}, it was proposed to only consider Eq. (\ref{ge}) to estimate the contribution to the HHG signal from each occupied MO. The contribution in Eq.(\ref{ge}) provides a qualitative trend, and allows one to balance accuracy and computational effort. However, the importance of each contribution in the decomposition depends on the system studied and must be carefully considered.

Using Eq.~(\ref{eq:spectrum}), we arrive at a MO-decomposition of the HHG spectrum
\begin{equation}
P(\omega) = \sum_i^\text{occ} |p_i(\omega)|^2 +  2\sum_i^\text{occ} \sum_{j>i}^\text{occ} p_i^*(\omega)p_j(\omega),
\label{}
\end{equation} 
where
\begin{equation}
p_i(\omega) = \frac{1}{t_\text{f} - t_\text{i}}\int^{t_\text{f}}_{t_\text{i}} W(t) {\boldsymbol  \mu}_i(t) \cdot {\bf n}_{\alpha'} \;  e^{-i\omega t} dt.
\label{eq:mos}
\end{equation} 
The contribution to the HHG spectrum coming only from the $i$-th MO can be defined as
\begin{equation}
P_i(\omega) = \vert p_i(\omega) \vert^{2},
\label{eq:mos}
\end{equation}
while the interference contribution between the $i$-th and $j$-th MOs can be measured by the relative phase $\Phi_{i,j}$
\begin{equation}
\text{cos}(\Phi_{i,j}) = \text{Re}\left(\frac{{p}^*_i(\omega) {p}_j(\omega)}{\vert {p}_i(\omega) \vert \vert {p}_j(\omega)\vert}\right),
\label{interference}
\end{equation}
and magnitude $M_{i,j}$
\begin{equation}
M_{i,j} = \vert p_i(\omega) \vert \vert p_j(\omega) \vert.
\label{magn}
\end{equation}
Furthermore, it is also interesting to measure the relative importance of the ground, ground-excited, and excited-excited contributions to $P_i(\omega)$ by defining
\begin{equation}
P_{i,\Lambda}(\omega) = \bigg\vert \frac{1}{t_\text{f} - t_\text{i}}\int^{t_\text{f}}_{t_\text{i}} W(t) \boldsymbol{\mu}_{i,\Lambda}(t) \cdot {\bf n}_{\alpha'} \; e^{-i\omega t} dt \bigg\vert^{2}, 
\label{PiLambdaomega}
\end{equation} 
where $\Lambda\in\{ \text{G},\text{G-E},\text{E-E}\}$.

\section{Computational Details}
\label{det}

In the RT-TD-CIS and RT-TD-CIS-LC-$\omega$PBE methods, the excited-state energies and transition-dipole moments are obtained from calculations performed with CIS and linear-response TD-LC-$\omega$PBE ($\omega=0.4$ $a_0^{-1}$) with the TDA \cite{pau21}, respectively, using the Q-Chem software package \cite{Shao:2006kl}. Then, they are employed in the Light code~\cite{lupp+12mol,lupp+13jcp,white15,coccia16a,coccia16b} which propagates the time-dependent wave function under the influence of a time-dependent electric field (see Section \ref{theo} for details). 

These electron dynamics are performed at fixed nuclear geometries. For CO$_2$, we use the experimental equilibrium distance of 2.209 $a_0$,  while for H$_2$O we optimize the geometry at the DFT level using the PBE functional, obtaining an equilibrium distance of 1.831 $a_0$ and a bond angle of 104.2$^{\circ}$.

The HHG spectra are calculated according to the computational strategy that we developed in recent years, which demonstrated to be successful for atoms and molecules.~\cite{coccia16b,ec19b,ec20,la18,white15,lup21,ec21}. We use augmented Gaussian basis sets, which are able to describe Rydberg  and low-energy continuum states. The accuracy of a Gaussian representation of HHG spectra has been extensively studied in literature. \cite{kauf+89physb,lupp+13jcp,white15,coccia16a,coccia16b,ec19b,mor23,woz21,woz24,woz24a} 
For CO$_{2}$, we combine a 5-fold augmented Dunning \cite{Dun-JCP-89} 5aug-cc-pVDZ basis set with 5 optimized Gaussian continuum functions (K) \cite{kauf+89physb}, obtaining the basis set named 5aug-cc-pVDZ+5K. For H$_{2}$O, we combine a 5-fold augmented Dunning 5aug-cc-pVTZ basis set with 5 optimized Gaussian continuum functions (K) obtaining a basis set named 5aug-cc-pVTZ+5K basis set. In both cases, 800 electronic excited states have been used for the expansion of the time-dependent wave function of Eq.~(\ref{wft}). This choice allows us to include all the ionization channels studied in this work. Moreover, this expansion of the time-dependent wave function represents a reliable balance between accuracy and computational effort. Spectra up to harmonic H31 (H35) are reported for CO$_2$ (H$_2$O), according to the maximum excitation energy contained in the time-dependent wave function (the notation H$M$ refers to the $M$-th harmonics in the spectrum).

We compute HHG spectra for a cos$^2$-shaped laser field [see Eq.~(\ref{coslaser})] with carrier frequency $\omega_0$=0.057 Ha (1.55 eV, 800 nm) and intensity $I= {\cal E}_{0}^{2}/2 = 8.5\times 10^{13}$ W/cm$^2$. The duration of the pulse is 23 optical cycles (oc): $\sigma$ = 23 oc [see Eq. (\ref{coslaser})], where 1 oc = 2$\pi$/$\omega_0$. The time step is 1.21 as (0.05 au). 

To prevent the nonphysical reflections of the wave function in the laser-driven electron dynamics, we use the  heuristic lifetime model (HLM) originally proposed Klinkusch {\it et al.}. \cite{Klinkusch:2009iw,coccia16b,ec19b,ec20,la18}  The HLM consists of adding to the energies $E_n$ calculated from the field-free Hamiltonian an imaginary term $-i \Gamma_n/2$ where $\Gamma_n = \sum_{i}^\text{occ} \sum_{a}^\text{vir} \theta(\epsilon_a) \vert r^a_{i,n}\vert^2 \sqrt{2\epsilon_a}/d$ (where $\theta$ is the Heaviside step function) represents the inverse lifetime for the excited state $n$ above the ionization threshold.
The parameter $d$ is empirically chosen, representing the characteristic escape length that the electron is allowed to travel during the lifetime $1/\Gamma_n$. This parameter was evaluated on the basis of the 3SM, by taking $d$ equal to the maximum electron excursion after ionization, i.e.
 15.143 $a_{0}$ in the present conditions. \\
In this work, by default, HHG spectra are calculated along the laser-pulse polarization direction, unless explicitly stated otherwise.

\section{Results and Discussion}
\label{res}

\subsection{CO$_2$}

The CO$_2$ ionization channels involved in the electron dynamics induced by the electric field considered here are $1^2\Pi_g$ (channel {\bf X}), $1^2\Pi_u$ (channel {\bf A}), $1^2\Sigma_u^+$ (channel {\bf B}), and $1^2\Sigma_g^+$ (channel {\bf C}). \cite{rub18,Smirnova2009} In Tab.~\ref{ips} and \ref{diffips} the experimental ionization energies of the  different channels, and  the corresponding differences in energy between these channels, are reported. The channel {\bf X} corresponds to the ground state of the CO$_2$ cation, and it shows the lowest ionization energy. Channels {\bf A} and {\bf B} are very close in energy (0.8 eV) but their symmetries are different: {\bf A} is $1^2\Pi_u$ and {\bf B} is $1^2\Sigma_u^+$. Channel {\bf C} is about 5.6 eV higher than channel {\bf X} and therefore energetically less accessible. 

\begin{figure}[!ht]
\includegraphics[width=0.8\textwidth]{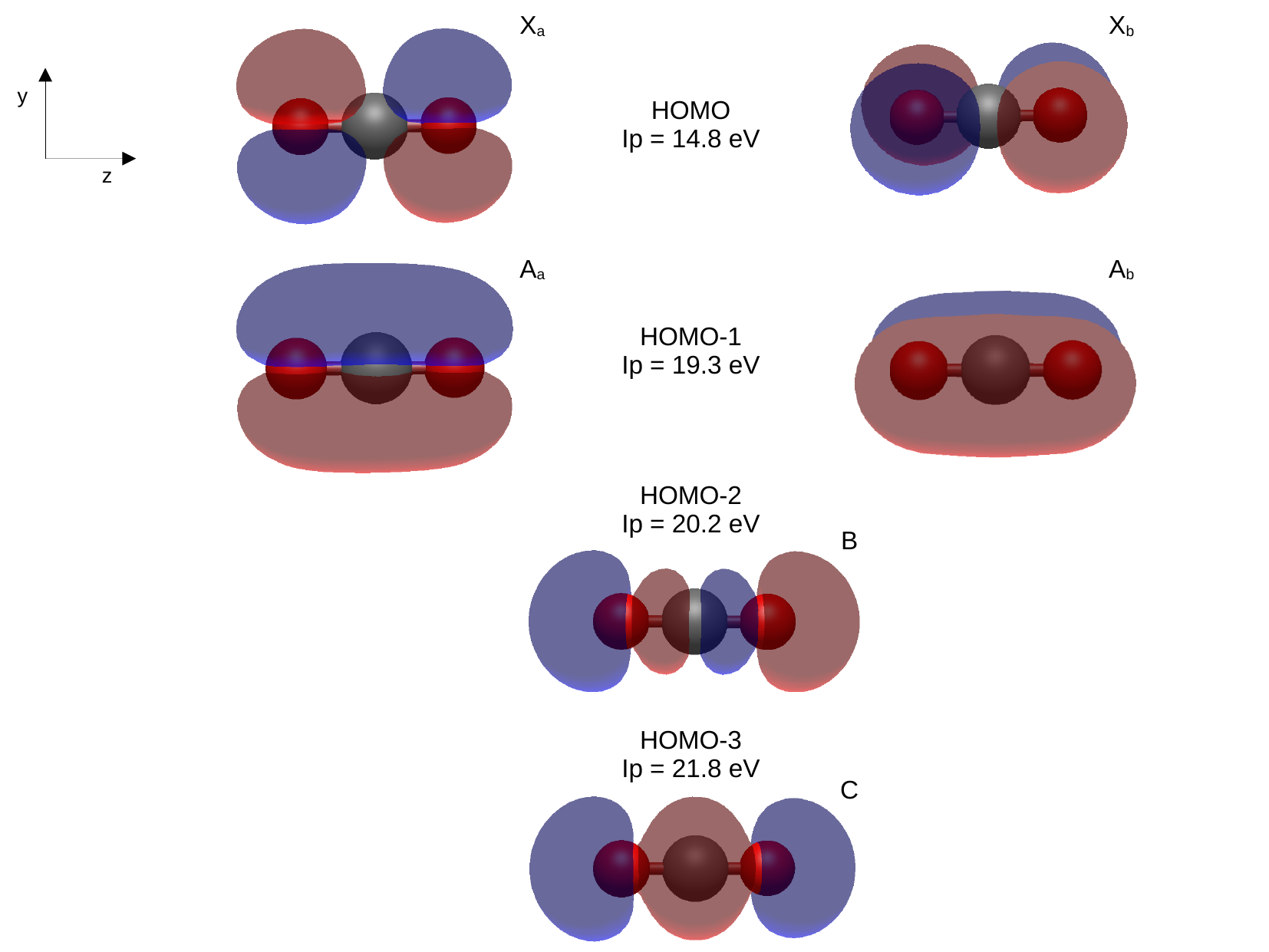}
\caption{\label{OrbitalsCIS} {HF occupied orbitals of CO$_{2}$.}}
\end{figure}

In Tab.~\ref{ips} we also report the theoretical energies of the ionization channels at the level of the HF theory using Koopmans' theorem. The HF occupied orbitals are  approximations to the ionization Dyson orbitals \cite{OrtizMP2019,5.0016472} which are overlaps between states with $N$ and $N-1$ electrons, corresponding to an electron that is removed from the molecule. \cite{rub18} The HF occupied orbitals are shown in Fig.~\ref{OrbitalsCIS}.
The HF ionization channels in Tab.~\ref{ips} correspond to the doubly-degenerate HOMO (1$\pi_g$, channel  {\bf X}: {\bf X}$_a$ and {\bf X}$_b$), the doubly-degenerate HOMO-1 (1$\pi_u$, channel {\bf A}: {\bf A}$_a$ and {\bf A}$_b$),  HOMO-2 (2$\sigma_u$, channel {\bf B}) and HOMO-3 (2$\sigma_g$, channel {\bf C}).  The corresponding ionization energy differences are reported in Tab.~\ref{diffips}. Also in this case, the energy difference between the {\bf A} and {\bf B} channels is 0.8 eV. The energy between the {\bf X} and {\bf C} channels is 7.047 eV which is 1.447 eV larger than the experimental value. Similar values have also been found in Ref. \citenum{rub18}. 

In calculating the HHG spectrum contributions for the different MOs, we took into account the orbital degeneracy by averaging the time-dependent dipole moment over degenerate orbitals and then Fourier transforming it.
The CO$_2$ molecule is aligned along the $z$ axis, as shown in Fig. \ref{OrbitalsCIS}.
In Fig. S1 of the Supporting Information (SI) we show the time-dependent dipole moment for a laser field polarized along $z$ of {\bf X}$_a$ and {\bf X}$_b$ (degenerate orbitals of {\bf X}), of {\bf A}$_a$ and {\bf A}$_b$ (degenerate orbitals of {\bf A}), of {\bf B} and of {\bf C}. In Fig. S2 of the SI, we show the same but for the time-dependent dipole moment in the case of a laser field polarized along $y$, i.e. perpendicular to the molecular axis.

%%%%%%%%%%%%%%%%%%%%%%%%%%%%%%%%%%%%%%%%%%%%%%%%%%%%%%%%%%%%%%%%%%%%%%%%%%%%%%%%%%%%%%%%%%%%%%%%
\begin{table}[!t]
\begin{center}
\begin{tabular}{ ccccc } 
\hline\hline
& Exp. (eV) \cite{kim81} & & & HF (eV)   \\
\hline
\vspace{0.1cm}
$1^2\Pi_g$ {\bf X} &  13.8 & 1$\pi_g$& HOMO & 14.8 \\
\vspace{0.1cm}
$1^2\Pi_u$ {\bf A}  &  17.3 &1$\pi_u$& HOMO-1 & 19.3\\
\vspace{0.1cm}
$1^2\Sigma_u^+$ {\bf B} &  18.1 &2$\sigma_u$& HOMO-2 & 20.2 \\
\vspace{0.1cm}
$1^2\Sigma_g^+$ {\bf C} &  19.4 &2$\sigma_g$& HOMO-3 & 21.8 \\
\hline\hline
\end{tabular}
\end{center}
\caption{Ionization energies (eV) for the channels of CO$_2$.}
\label{ips}
\end{table}
%%%%%%%%%%%%%%%%%%%%%%%%%%%%%%%%%%%%%%%%%%%%%%%%%%%%%%%%%%%%%%%%%%%%%%%%%%%%%%%%%%%%%%%%%%%%%%%%

%%%%%%%%%%%%%%%%%%%%%%%%%%%%%%%%%%%%%%%%%%%%%%%%%%%%%%%%%%%%%%%%%%%%%%%%%%%%%%%%%%%%%%%%%%%%%%%%
\begin{table}[!t]
\begin{center}
\begin{tabular}{ cccc } 
\hline\hline
& $\Delta$Exp. (eV) &  $\Delta$HF (eV)   \\
\hline
\vspace{0.1cm}
{\bf X}-{\bf A} &  3.5 & 4.5 \\
\vspace{0.1cm}
{\bf X}-{\bf B}  &  4.3 & 5.4\\
\vspace{0.1cm}
{\bf X}-{\bf C} &  5.6 & 7.0 \\
\vspace{0.1cm}
{\bf A}-{\bf B} &  0.8 & 0.9 \\
\vspace{0.1cm}
{\bf A}-{\bf C} &  2.1 & 2.5 \\
\vspace{0.1cm}
{\bf B}-{\bf C} &  1.3 & 1.6 \\
\hline\hline
\end{tabular}
\end{center}
\caption{Ionization energy differences (eV) between the channels of CO$_2$.}
\label{diffips}
\end{table}
%%%%%%%%%%%%%%%%%%%%%%%%%%%%%%%%%%%%%%%%%%%%%%%%%%%%%%%%%%%%%%%%%%%%%%%%%%%%%%%%%%%%%%%%%%%%%%%%

%%%%%%%%%%%%%%%%%%%%%%%%%%%%%%%%%%%%%%%%%%%%%%%%%%%%%%%%%%%%%%%%%%%%%%%%%%%%%%%%%%%%%%%%%%%%%%%%
%%%%%%%%%%%%%%%%%%%%%%%%%%%%%%%%%%%%%%%%%%%%%%%%%%%%%%%%%%%%%%%%%%%%%%%%%%%%%%%%%%%%%%%%%%%%%%%%

\begin{figure}[!ht]
\includegraphics[width=0.8\textwidth]{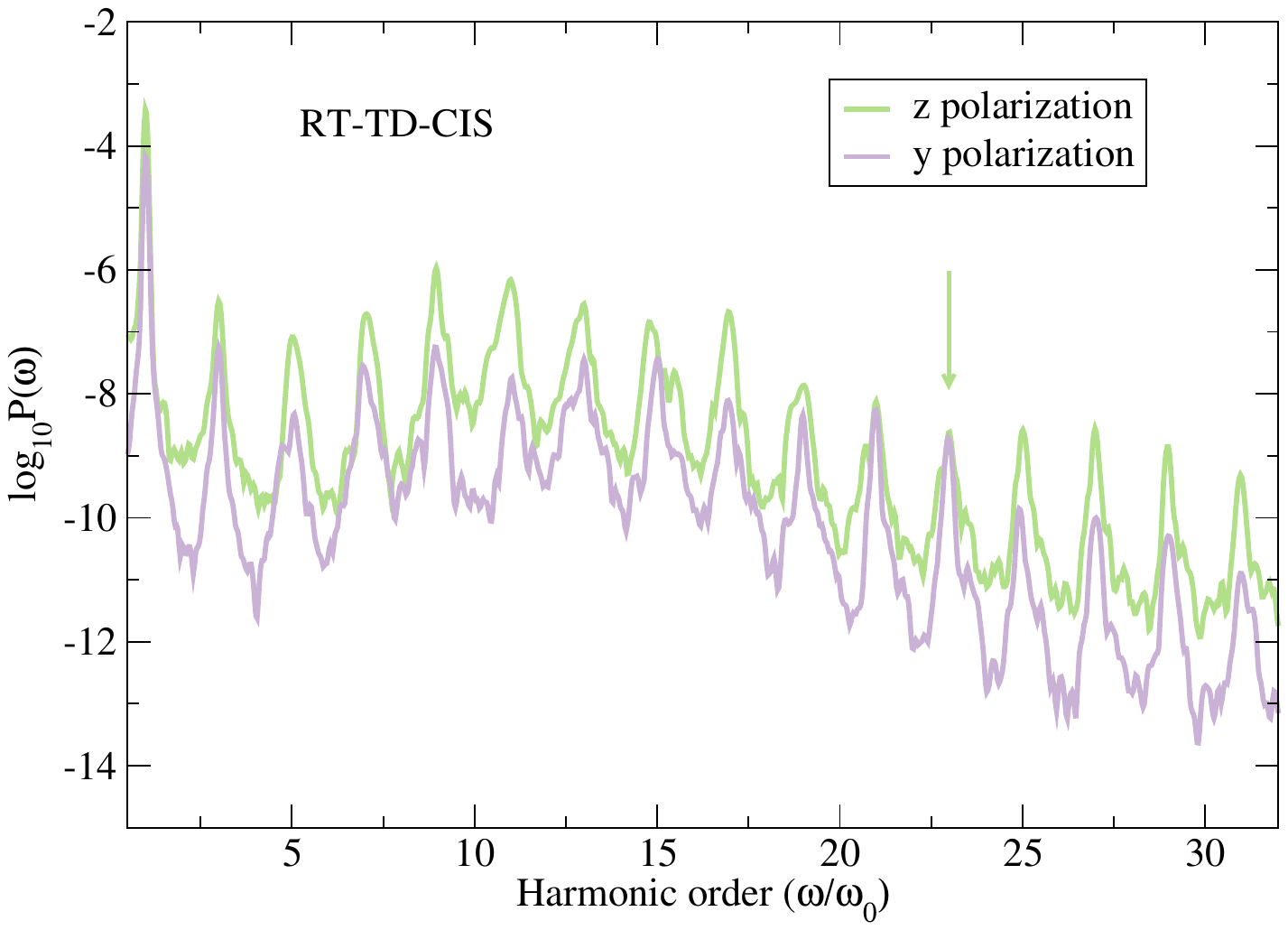}
\caption{\label{HHG_CIS} Total HHG spectra for CO$_{2}$, with laser-pulse polarization along $y$ axis (light purple line) and along $z$ axis (light green line), at the RT-TD-CIS level of theory. The arrow represents the minimum for the $z$-polarized case.}
\end{figure}

In Fig.~\ref{HHG_CIS} we report the HHG spectra of CO$_{2}$ computed with RT-TD-CIS when the laser field is polarized in the $z$  or $y$  direction. As CO$_{2}$ has inversion symmetry, only odd harmonics are generated. \cite{liu16} For the $z$ polarization, we observe the expected minimum at the harmonic 23 (H23, light green arrow). \cite{rub18,wor10} 

%%%%%%%%%%%%%%%%%%%%%%%%%%%%%%%%%%%%%%%%%%%%%%%%%%%%%%%%%%%%%%%%%%%%%%%%%%%%%%%%%%%%%%%%%%%%%%%%
%%%%%%%%%%%%%%%%%%%%%%%%%%%%%%%%%%%%%%%%%%%%%%%%%%%%%%%%%%%%%%%%%%%%%%%%%%%%%%%%%%%%%%%%%%%%%%%%

%%%%%%%%%%%%%%%%%%%%%%%%%%%%%%%%%%%%%%%%%%%%%%%%%%%%%%%%%%%%%%%%%%%%%%%%%%%%%%%%%%%%%%%%%%%%%%%%
\begin{figure}[!ht]
\includegraphics[width=0.8\textwidth]{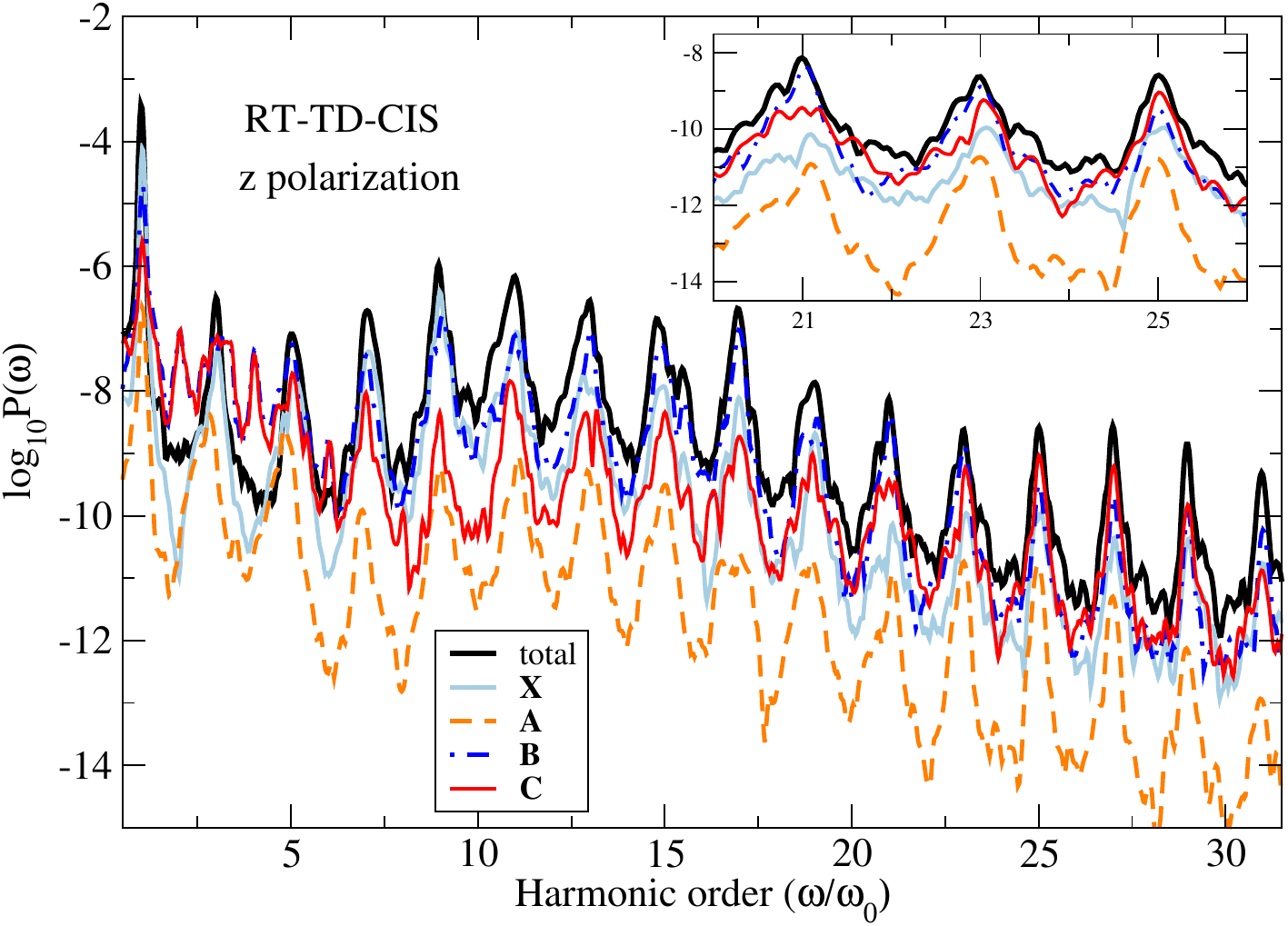}
\includegraphics[width=0.8\textwidth]{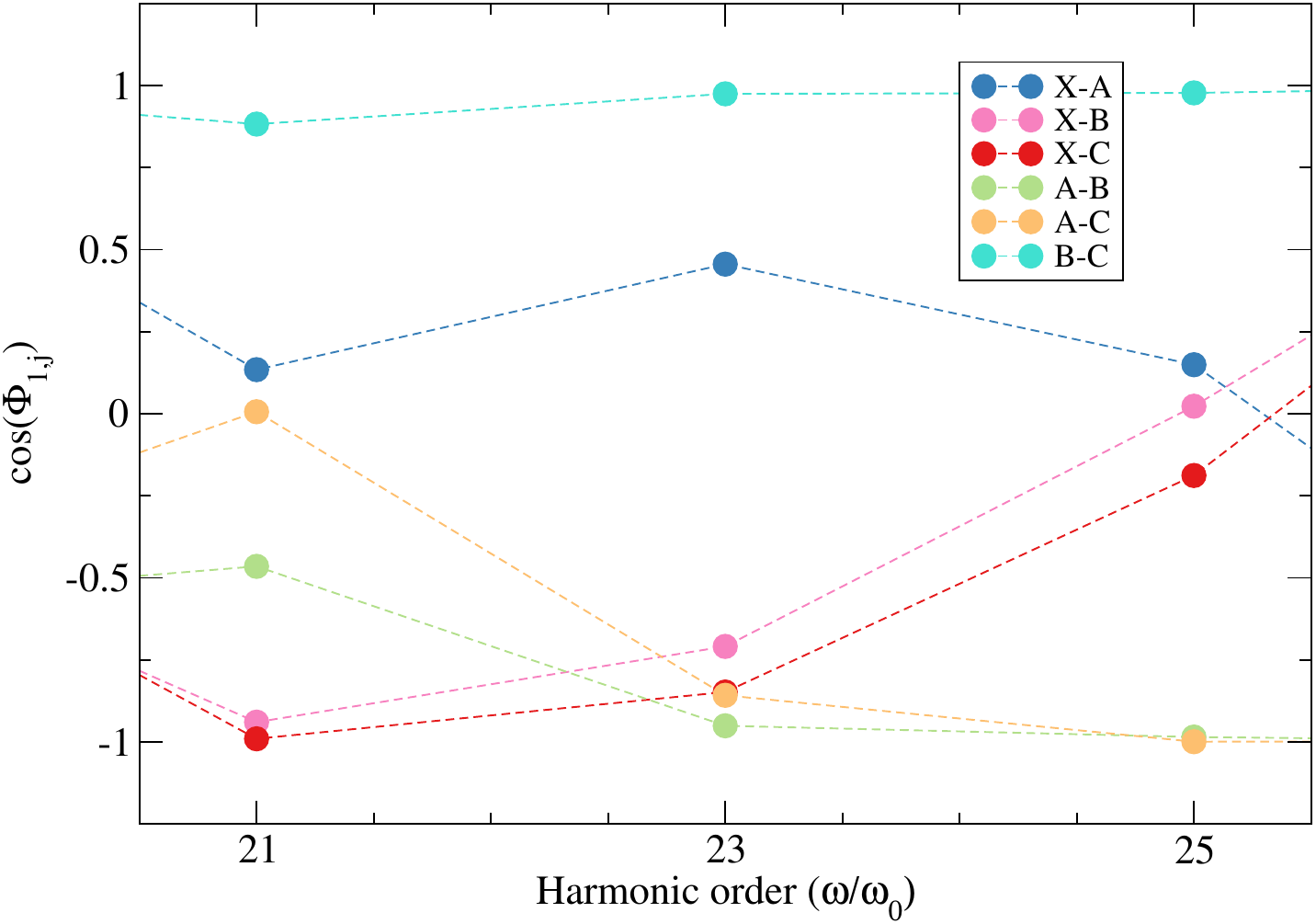}
\caption{\label{HHG_CIS_Z_MO} Top: MO decomposition of the HHG spectrum of CO$_2$, with laser-pulse polarization along the $z$ axis, at the RT-TD-CIS level of theory. The inlet plot is a zoom on the minimum region (H21-H25). Bottom: Interference contributions between different ionization channels with laser-pulse polarization along the $z$ axis, at the RT-TD-CIS level of theory.}
\end{figure}
%%%%%%%%%%%%%%%%%%%%%%%%%%%%%%%%%%%%%%%%%%%%%%%%%%%%%%%%%%%%%%%%%%%%%%%%%%%%%%%%%%%%%%%%%%%%%%%%

%%%%%%%%%%%%%%%%%%%%%%%%%%%%%%%%%%%%%%%%%%%%%%%%%%%%%%%%%%%%%%%%%%%%%%%%%%%%%%%%%%%%%
\begin{figure}[!ht]
\includegraphics[width=0.8\textwidth]{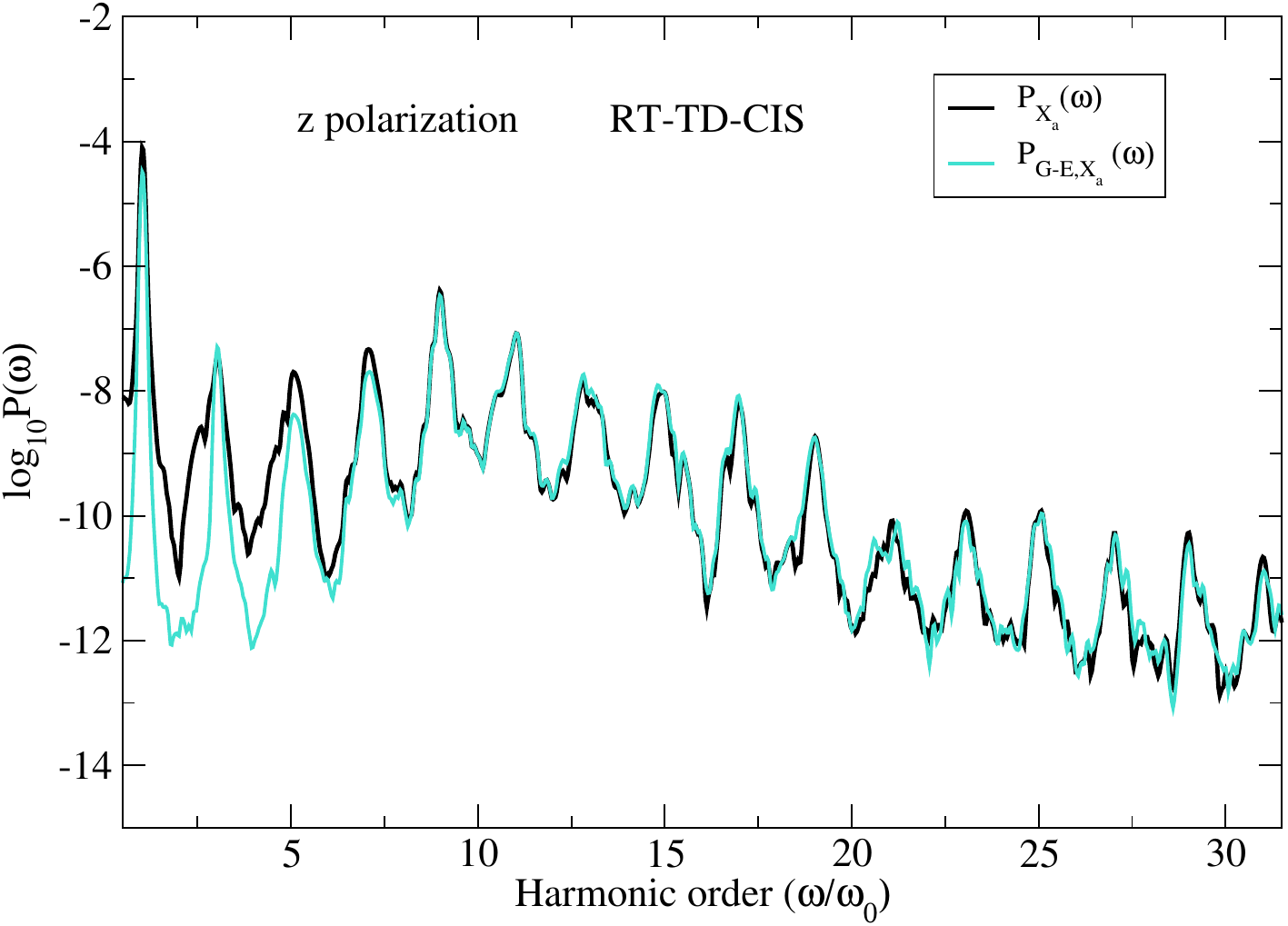}
\includegraphics[width=0.8\textwidth]{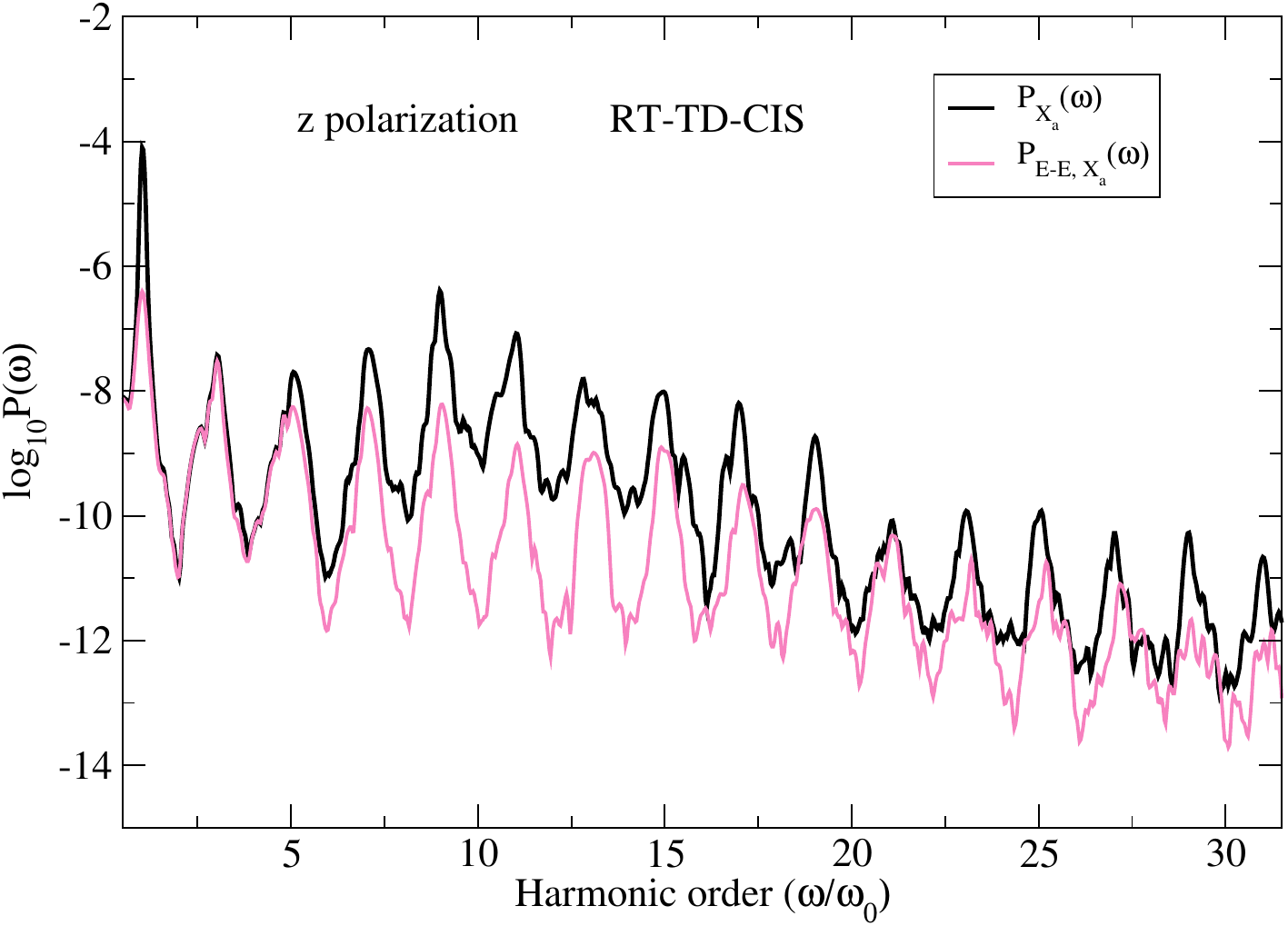}
\caption{\label{HHG_comp}Ground-excited (G-E, top) and excited-excited (E-E, bottom) contributions to the HHG spectrum for the ${\bf X}_a$ channel of CO$_2$, with laser-pulse polarization along the $z$ axis, at the RT-TD-CIS level of theory.}
\end{figure}
%%%%%%%%%%%%%%%%%%%%%%%%%%%%%%%%%%%%%%%%%%%%%%%%%%%%%%%%%%%%%%%%%%%%%%%%%%%%%%%%%%%%%

In Fig.~\ref{HHG_CIS_Z_MO} we show the MO contributions  (top panel) and the interference contributions (bottom panel) to the HHG spectrum for laser field polarized along the $z$ direction. As a comparison, we also report the total HHG spectrum.

The interpretation of the MO contributions relies on two aspects: symmetry and energy of each orbital. \cite{PhysRevA.71.061801} Spatial symmetry and ionization energy are the two essential factors to be accounted for describing the strong-field electron dynamics generating the HHG spectrum. In fact, ionization and recombination processes depend on the energy accessibility of a particular MO and also on the particular symmetry of the MO along the polarization direction of the laser field. 
Thanks to our approach, we are able to see which part of the HHG spectrum is controlled by a given channel, and by interferences between them. Indeed, the contribution of a given MO is energy-dependent, and it can largely change in the spectral window of interest.

The HOMO ({\bf X} channel) highly contributes because of the lowest ionization energy. However, the HOMO-2 ({\bf B} channel), which is less energetically favorable than {\bf X}, contributes almost equally the {\bf X} channel, becoming the leading contribution from H11 to H25. This is due to the symmetry of the HOMO-2  which favors ionization along the axis of the molecule. In fact, HOMO-2 is a $\sigma_u$ orbital, and it has maximal electron density along the direction of the internuclear axis. HOMO-3 ({\bf C} channel) produces a non-negligible HHG spectrum, especially at large energies, where it gives the largest contribution for H27-H31. Also in this case, it is the symmetry of the orbital $\sigma_g$ that plays an important role. In fact, from the energetic point of view the HOMO-3 is not favorable, as it is 7.047 eV lower than the HOMO. However, its symmetry makes it as important as the HOMO-2 which is only 5.405 eV lower than HOMO. HOMO-1 (channel {\bf A}) would be in principle energetically most favorable
than HOMO-2 and HOMO-3 for ionization, however as clearly observed in Fig.~\ref{HHG_CIS_Z_MO}, its contribution is much smaller than those of the other channels. HOMO-1 has an unfavorable $\pi_u$ symmetry along the $z$ direction. 
The same general shape of the full spectrum, with the minimum at H23, and the same partial contributions from individual MOs were observed when the HHG spectra were computed in the velocity and acceleration forms, as described in the Appendix of Ref. \citenum{coccia16b}  (Figure S3 in the SI).

In the bottom panel of Fig.~\ref{HHG_CIS_Z_MO} we show the interference contributions between the different channels of CO$_2$ [see Eq.~(\ref{interference})]. The interference contributions between channels are supposed to be responsible for the dynamical minimum in the HHG spectrum, as reported in Ref. \citenum{Smirnova2009,wor10}. In particular, the minimum is attributed to destructive interferences, i.e. cos($\Phi_{ij})<0$ (Eq. \ref{interference}) between the ({\bf X} channel) HOMO and ({\bf B} channel) HOMO-2 (X-B label), \cite{Smirnova2009} as observed in Fig. \ref{HHG_CIS_Z_MO}. Indeed, the interference for the {\bf X}-{\bf B} channels shows the largest magnitude [Eq.~(\ref{magn})] among the computed negative-phase interferences, as reported in Fig. S4 of the SI.

However, the analysis of the interferences in Fig.~\ref{HHG_CIS_Z_MO} also points out the complexity of the electron dynamics.
In fact, considering that the minimum corresponds to H23, we see that going from H21 to H25 there is a phase change from positive to negative for other channels as well. 
Observing a dynamical minimum thus seems to be related to multiple interference contributions.

Ruberti {\it et al.} \citenum{rub18} analyzed the multi-channel dynamics of aligned CO$_2$ with B-splines time-dependent first-order algebraic-diagrammatic-construction calculations (TD-ADC(1)). The methodology we propose here gives the same trend as TD-ADC(1) for the single MO contributions to the HHG spectrum.  Moreover, also in the TD-ADC(1) calculations, the minimum mainly corresponds to the destructive interference between the HOMO and the HOMO-2 channels.

\begin{figure}[!ht]
\includegraphics[width=0.8\textwidth]{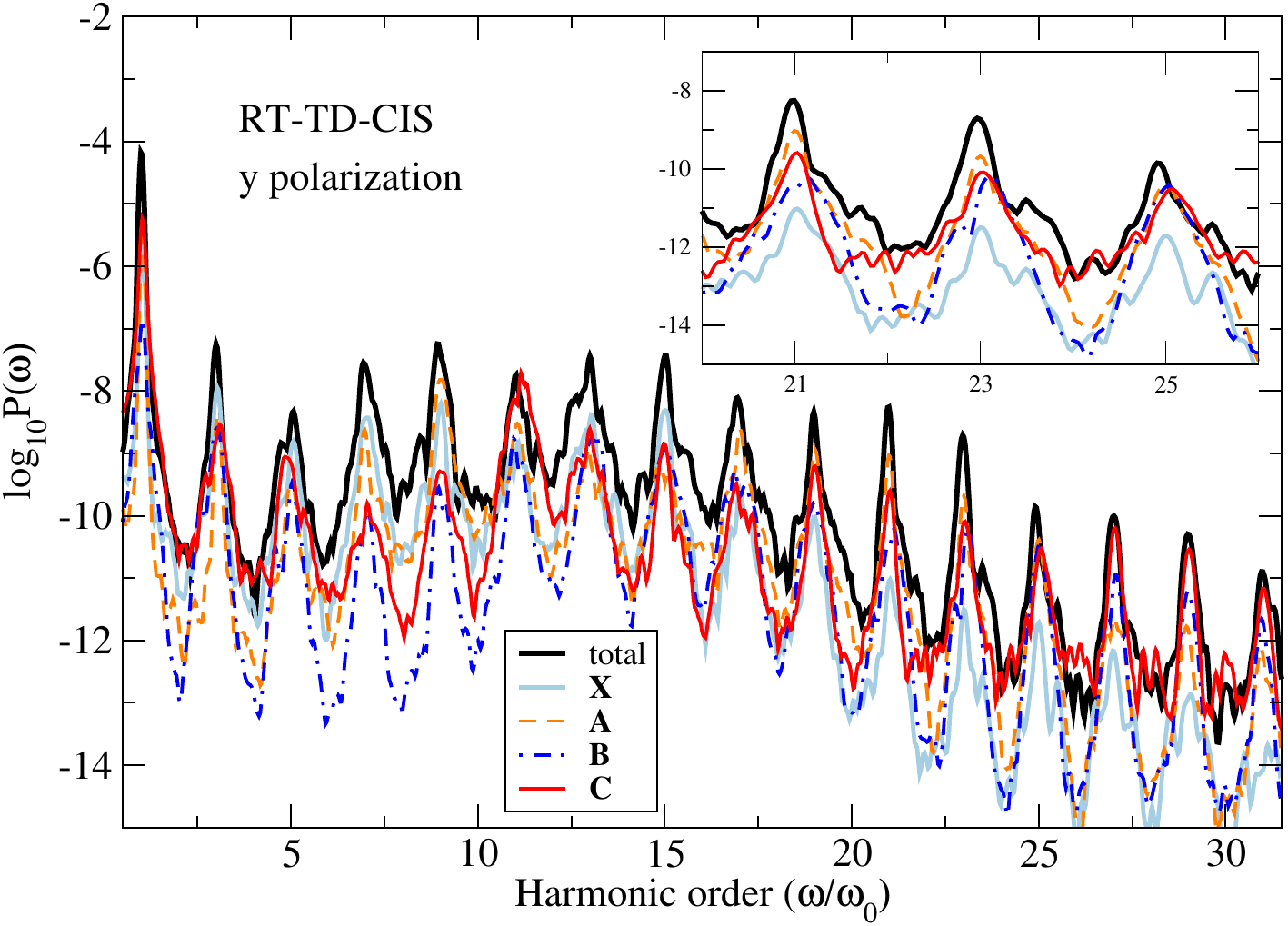}
\includegraphics[width=0.8\textwidth]{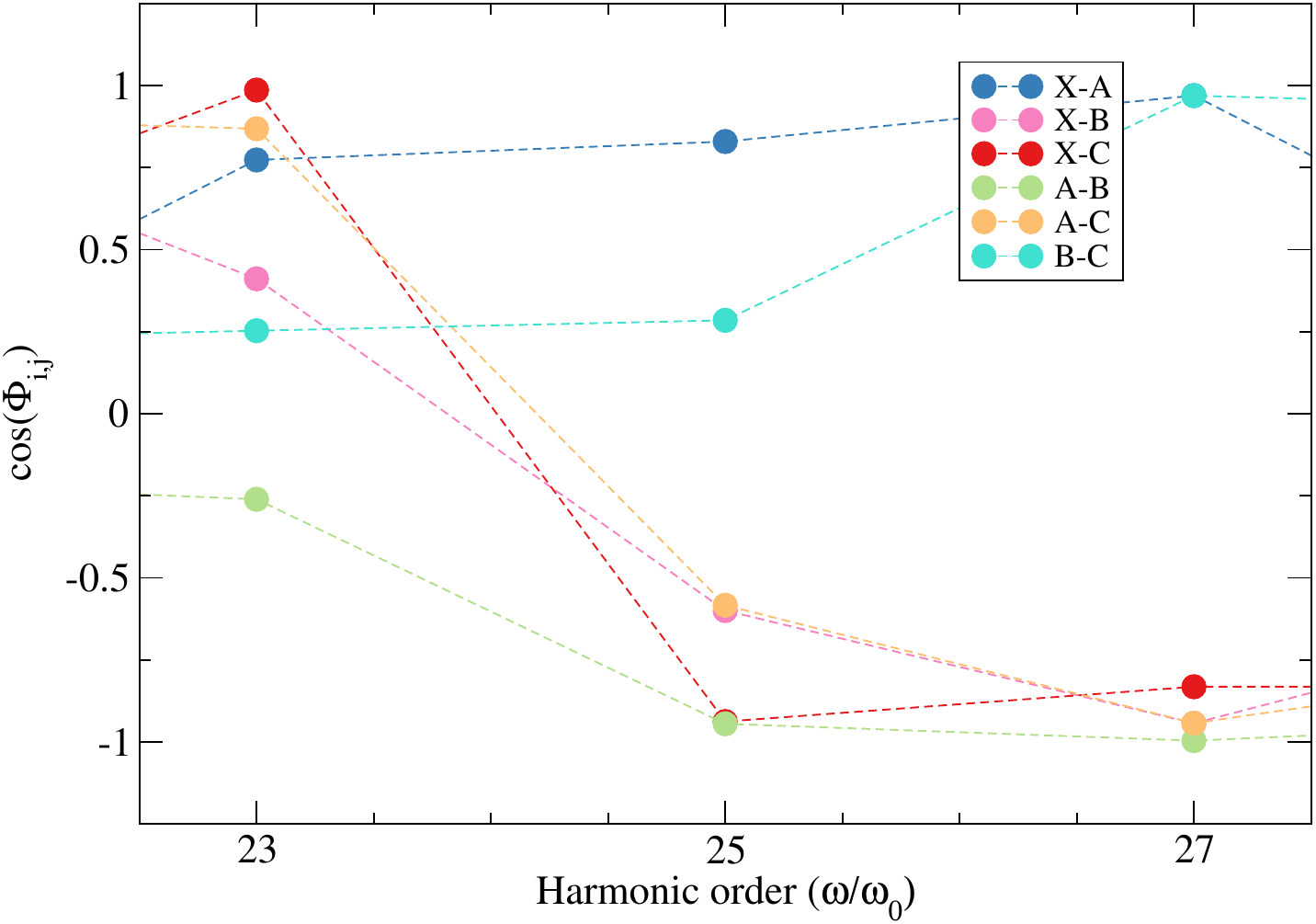}
\caption{\label{HHG_CIS_Y_MO} Top: MO decomposition of the HHG spectrum of CO$_2$, with laser-pulse polarization along the $y$ axis, at the RT-TD-CIS level of theory. The inlet plot is a zoom on the region H21-H25. Bottom: Interference contributions between different ionization channels with laser-pulse polarization along the $y$ direction, at the RT-TD-CIS level of theory. }
\end{figure}

The MO analysis of the HHG spectrum can be refined using the ground-excited and excited-excited contributions in Eq.~(\ref{PiLambdaomega}). In Fig.~\ref{HHG_comp} we show for the channel {\bf X} that 
the contribution $P_{i,\text{G-E}}(\omega)$ is almost identical to the full spectrum $P_{i}(\omega)$ for this orbital. 
The corresponding contributions to the dipole moment are collected in Fig. S5 of the SI.

The advantage to approximate the MO contribution to the HHG spectrum from only the contribution $\boldsymbol{\mu}_{i,\text{G-E}}(t)$ to the dipole moment is that more complex systems with large basis sets (i.e. computationally expensive) can be calculated. We note that the present study confirms that it is indeed reasonable to approximate the MO decomposition of the HHG spectrum from only the contribution $\boldsymbol{\mu}_{i,\text{G-E}}(t)$ (possibly, adding also the ground-state contribution $\boldsymbol{\mu}_{i,\text{G}}(t)$ if it is not zero), as recently done by some of us in Refs.~\citenum{lup23,mor24}.

In Fig.~\ref{HHG_CIS_Y_MO} we report the MO contributions to the HHG spectrum (top panel) and the interference contributions (bottom panel) for a laser field polarized along the $y$ direction. We also report as a comparison the total HHG spectrum.  Also in this case, we observe the role of symmetry and ionization energy in the contribution of the MOs to the HHG spectrum. The {\bf X} channel, while not having a particularly favorable symmetry for the $y$ polarization, remains the most accessible in energy, and it is particularly important for the harmonics up to H15. The channels {\bf A} and {\bf B} are very close in energy but with different symmetries. The symmetry $\pi_u$ of the channel {\bf A} (HOMO-1)  is favored over that of the channel B ($\sigma_u$) for a laser polarized along the $y$ axis. This is observed in Fig.~\ref{HHG_CIS_Y_MO} where the channel {\bf A} (HOMO-1)  has a leading contribution to the whole spectrum, while the  channel {\bf B} (HOMO-2) contribution is much lower. Only for the highest harmonics, the channels {\bf A} and {\bf B} have the same order of magnitude in the HHG spectrum. The channel {\bf C} has higher ionization energy  than the {\bf A} and {\bf B} channels and is therefore more difficult to access. However, its $\sigma_g$ symmetry makes it competitive with both channels. The channel {\bf C} contribution is larger than {\bf B} and of the same order of magnitude as {\bf A}. 

In the bottom panel of Fig.~\ref{HHG_CIS_Y_MO} we show the interference contributions between the different channels of CO$_2$ for the $y$ polarization of the laser field [see Eq.~(\ref{interference})]. In this case, we focused on the H25 harmonic because there is a  significant change in the intensity with respect to the H23 harmonic. We therefore suspect that there are destructive interferences in the spectrum. In fact, we observe that going from H23 to H25 many phases change from positive to negative values. 
Here, the interference contribution between the {\bf X} channel (HOMO) and {\bf A} channel (HOMO-1)) is always positive. Therefore, the destructive interference is due to channels {\bf X} (HOMO) and {\bf C} (HOMO-2) and to the other inner MOs.
Also in this case we observe that the dynamics is quite complicated and involves multiple-MO contributions.  

With our methodology, we also investigate the role of MOs in RT-TD-CIS-LC-$\omega$PBE. We have explicitly chosen this level of theory because an inversion in the order of the MOs is found. This means that in LC-$\omega$PBE calculations, the channel {\bf A} (HOMO-1) has $\sigma_u$ symmetry while channel {\bf B} (HOMO-2) has $\pi_u$ symmetry. This is shown in Figure S6 of the SI. Moreover, the corresponding LC-$\omega$PBE ionization energies are reported in Tab. S1 of the SI and their differences are in Tab. S2 of the SI. Regardless of the different order of the orbitals, the calculated total HHG spectrum (Fig. S7 of the SI) does not appear to be particularly different from that calculated in RT-TD-CIS, especially at lower harmonics, as observed in Fig. S8 and S9 of the SI. 

RT-TD-CIS-LC-$\omega$PBE was recently used by some of us to investigate the effect of HF exchange in the HHG spectra of H$_2$, N$_2$, and CO$_2$. By varying the range-separation parameter $\omega$, we transitioned from pure PBE ($\omega = 0$) to HF plus PBE correlation ($\omega \rightarrow \infty$). \cite{pau21} In this reference, differences in the computed spectra were observed in the HHG cutoff region: RT-TD-CIS and RT-TD-CIS-LC-$\omega$PBE with $\omega = 0.4$ (also used in this work) provided better resolution of the harmonics compared to RT-TD-CIS-PBE.

However, from the MO analysis (partial dipole moments are collected in Fig. S10 and S11 of the SI) we are able to infer information about the MO symmetry and to observe directly in the HHG spectrum the inverse ordering of the MOs. In the RT-TD-CIS-LC-$\omega$PBE calculations, the roles of the channel {\bf A} (HOMO-1) and  channel {\bf B} (HOMO-2) are exchanged with respect to RT-TD-CIS. This  is shown in Figs. S12 and S13 of the SI for laser-pulse polarization along the $z$ and $y$ axes, respectively.

\subsection{H$_2$O}
\label{h2o}

For H$_2$O, the experimental ionization energies of the different channels and the energy differences between them are reported in Tab.~\ref{tab:ip_h2o} and \ref{tab:endiffip_h2o}. We observe that all the MOs are energetically well separated ($>$ 1 eV). This was not the case for the HOMO-1 and the HOMO-2 in CO$_2$, where the energy difference was about 0.5 eV.

\begin{figure}[!ht]
\includegraphics[width=0.8\textwidth]{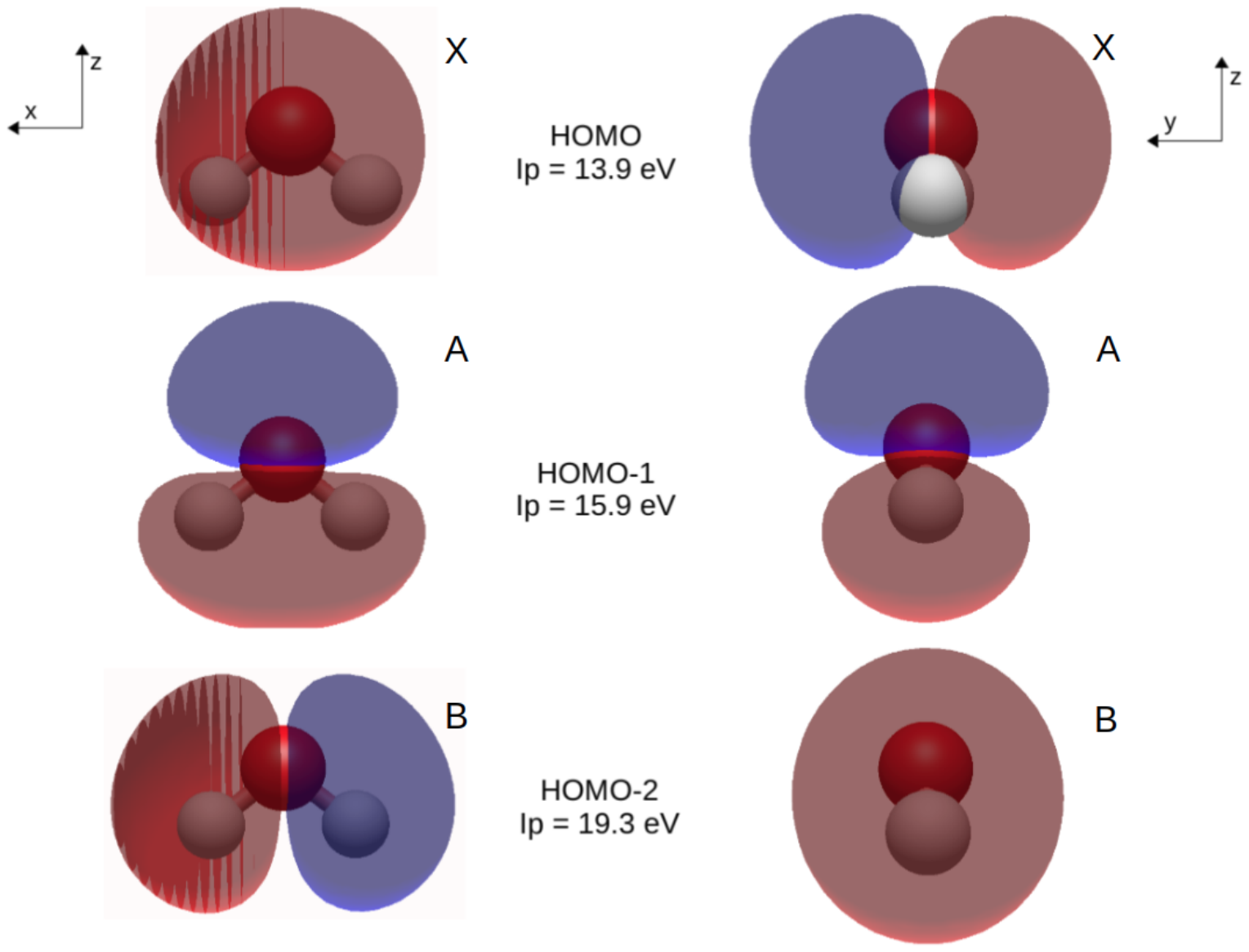}
\caption{\label{OrbitalsCISH2O} HF occupied orbitals of H$_2$O.}
\end{figure}

In Fig.~\ref{OrbitalsCISH2O}, we report the HF occupied orbitals of H$_2$O and their corresponding HF ionization energies. We observe that HOMO, HOMO-1, and HOMO-2 are p-type orbitals with a single nodal plane. 

%%%%%%%%%%%%%%%%%%%%%%%%%%%%%%%%%%%%%%%%%%%%%%%%%%%%%%%%%%%%%%%%%%
\begin{table}[!t]
\centering
\begin{tabular}{ccccc}
\hline\hline
&Exp.\cite{ning2008high}& & HF \\
\hline
$\tilde{\bf X}$ & 12.6  & 1b$_1$ & HOMO &  13.9\\
\vspace{1mm}
$\tilde{\bf A}$ &  14.8 & 2a$_1$ & HOMO-1 &  15.9 \\
\vspace{1mm}
$\tilde{\bf B}$ &  18.7  & 1b$_2$ & HOMO-2 & 19.3  \\   
\hline\hline   
\end{tabular}
\caption{Ionization energies (eV) for the channels of H$_2$O.}
\label{tab:ip_h2o}
\end{table}
%%%%%%%%%%%%%%%%%%%%%%%%%%%%%%%%%%%%%%%%%%%%%%%%%%%%%%%%%%%%%%%%%%

%%%%%%%%%%%%%%%%%%%%%%%%%%%%%%%%%%%%%%%%%%%%%%%%%%%%%%%%%%%%%%%%%%
\begin{table}[!t]
\centering
\begin{tabular}{ccccc}
\hline\hline
&$\Delta$Exp.& $\Delta$HF \\
\hline
$\tilde{\bf X}$-$\tilde{\bf A}$  & 2.2 & 2.0 \\
\vspace{1mm}
$\tilde{\bf X}$-$\tilde{\bf B}$  & 6.1 & 5.5\\
\vspace{1mm}
$\tilde{\bf A}$-$\tilde{\bf B}$  &3.9 & 3.4\\
\hline\hline
\end{tabular}
\caption{Ionization energy differences (eV) between the channels of H$_2$O.}
\label{tab:endiffip_h2o}
\end{table}
%%%%%%%%%%%%%%%%%%%%%%%%%%%%%%%%%%%%%%%%%%%%%%%%%%%%%%%%%%%%%%%%%%

Experimental literature on gas-phase H$_2$O reports HHG spectra under various laser-pulse intensities and wavelengths, showing an increased cutoff due to ionization suppression, when compared to the cutoff in the Xe atom, which has a similar ionization potential. \cite{wong2010high} Aligning water molecules would allow one to properly study MO contributions, but, to the best of our knowledge, such an experiment has been not carried out yet.
Recently, a theoretical approach for MO tomography of water HOMO and HOMO-1, based on RT-TDDFT, has been proposed. \cite{ren2023molecular} In that work, the multiple-MO nature of the H$_2$O HHG signal is observed. 

\begin{figure}[!ht]
\includegraphics[width=0.8\textwidth]{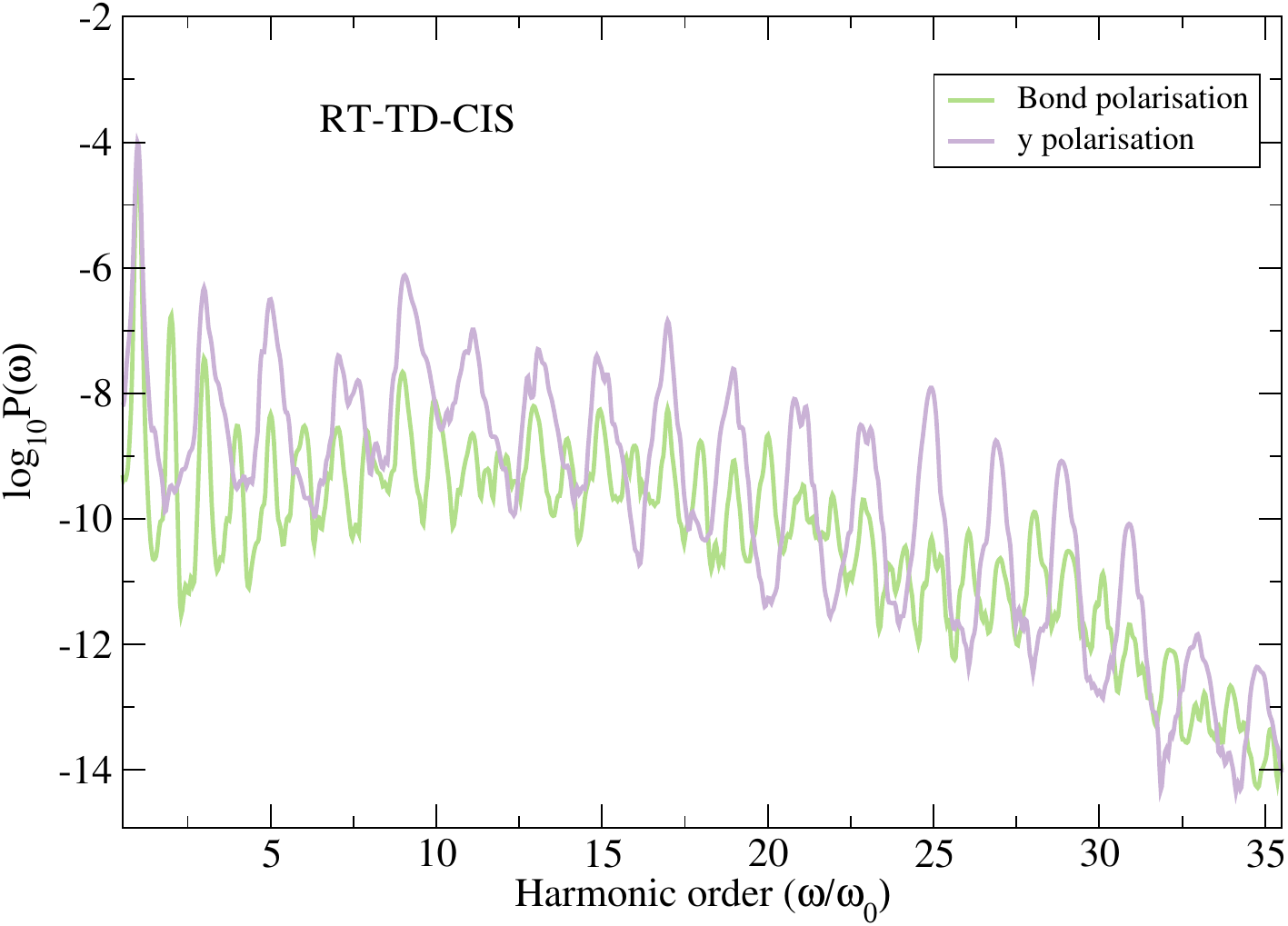}
\caption{\label{HHG_H2O} Total HHG spectra for H$_2$O, with laser-pulse polarization perpendicular to the molecular plane ($y$ polarization, light purple line) and parallel to a O-H bond (bond polarization, light green line), at the RT-TD-CIS level of theory.}
\end{figure}

We repeat for H$_2$O the same conceptual scheme applied to CO$_2$. The molecule is oriented in space with respect to the linearly polarized probe laser pulse, and we study the partial HHG signal from HOMO and inner MOs when the pulse direction is changed.
In Fig.~\ref{HHG_H2O} we show the total HHG spectra for the H$_2$O molecule when the laser is polarized along the $y$ direction, i.e. perpendicular to the plane ($xz$) of the molecule, and in the direction defined by a O-H bond. The spectra appear different both in terms of peak intensities and allowed harmonics. For the $y$ polarization, only odd harmonics are present, in contrast with the bond polarization which shows even and odd harmonics with lower intensity. 

When an aligned molecule is considered, the key condition to get even harmonics to vanish is the presence of a reflection plane perpendicular to the polarization direction. More generally, we can also interpret these results by considering the dynamical symmetry (DS) of the time-dependent Hamiltonian with a linearly polarized laser pulse along the $y$ axis \cite{neu19,tzu22} (see Fig.~\ref{OrbitalsCISH2O} for the reference system). Simply, the theory of DS includes the cases routinely known in HHG spectroscopy.
The reflection plane is not present with laser-pulse polarization along the O-H bond, which therefore provides odd and even harmonics.

%%%%%%%%%%%%%%%%%%%%%%%%%%%%%%%%%%%%%%%%%%%%%%%%%%%%%%%%%%%%%%%%%%%%%%%%%%%%
%%%%%%%%%%%%%%%%%%%%%%%%%%%%%%%%%%%%%%%%%%%%%%%%%%%%%%%%%%%%%%%%%%%%%%%%%%%%

%%%%%%%%%%%%%%%%%%%%%%%%%%%%%%%%%%%%%%%%%%%%%%%%%%%%%%%%%%%%%%%%%%%%%%%%%%%%
%%%%%%%%%%%%%%%%%%%%%%%%%%%%%%%%%%%%%%%%%%%%%%%%%%%%%%%%%%%%%%%%%%%%%%%%%%%%

\begin{figure}[!ht]
\includegraphics[width=0.8\textwidth]{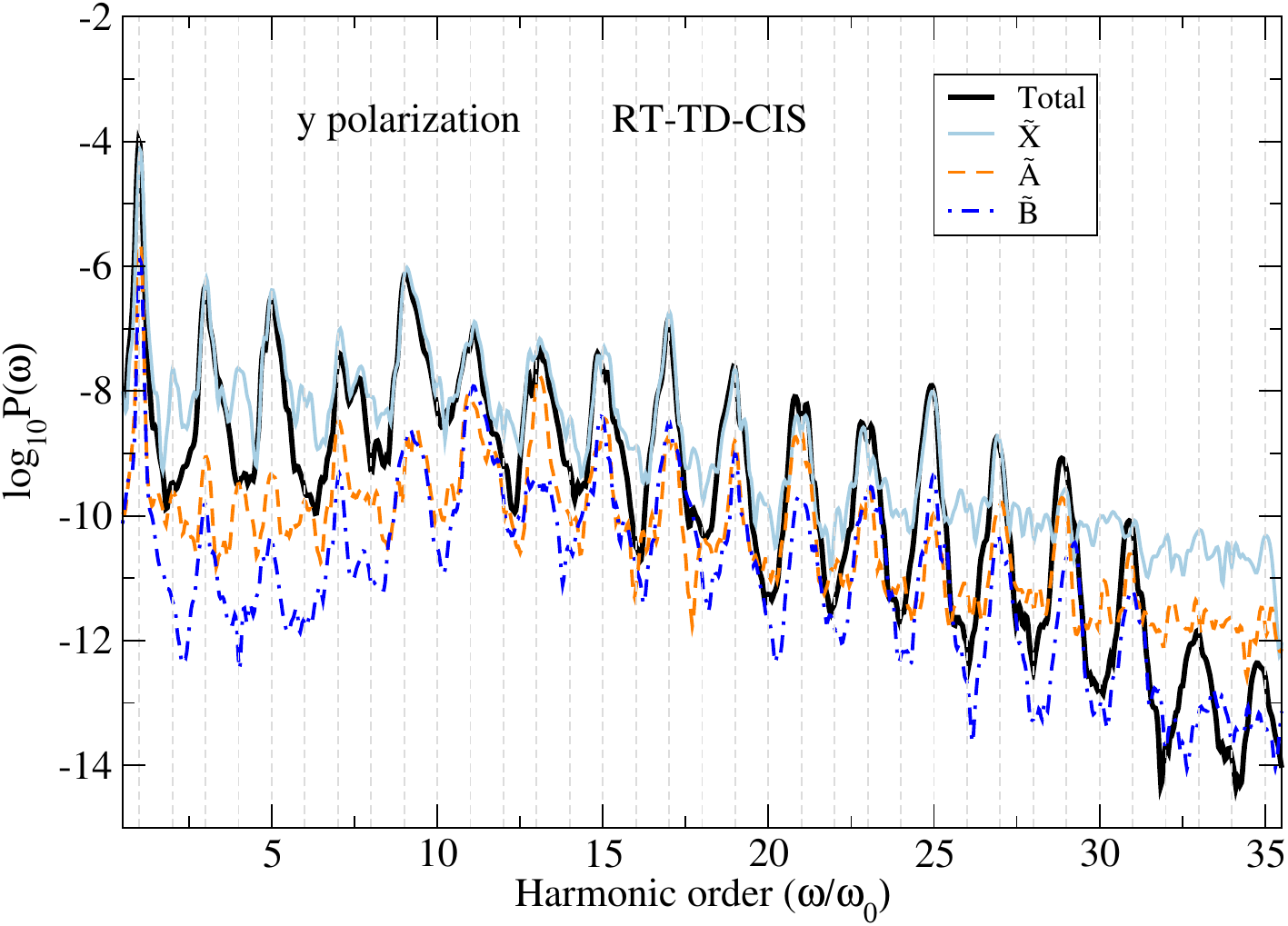}\\
 \includegraphics[width=0.8\textwidth]{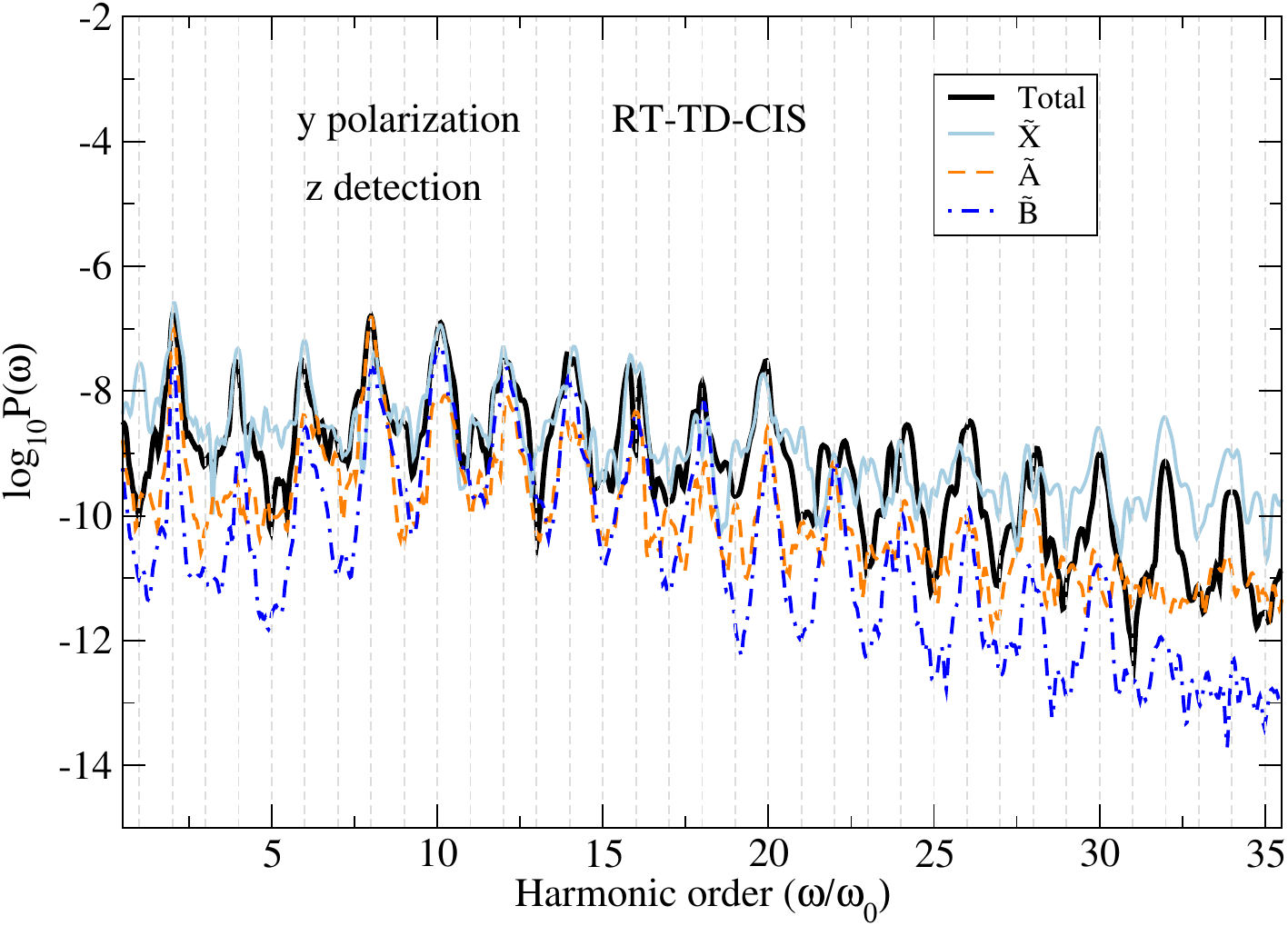}
\caption{\label{HHG_pol_Y_H2O}  HHG spectra of H$_2$O with laser-pulse polarization perpendicular to the molecular plane ($y$ direction) at the RT-TD-CIS level of theory. Top: Spectra are calculated along the laser-pulse polarization direction. Bottom: Spectra are calculated along the $z$-axis, orthogonal to the laser-pulse polarization direction. Vertical dashed lines are a guide for the eye.}
\end{figure}

\begin{figure}[!ht]
\includegraphics[width=0.8\textwidth]{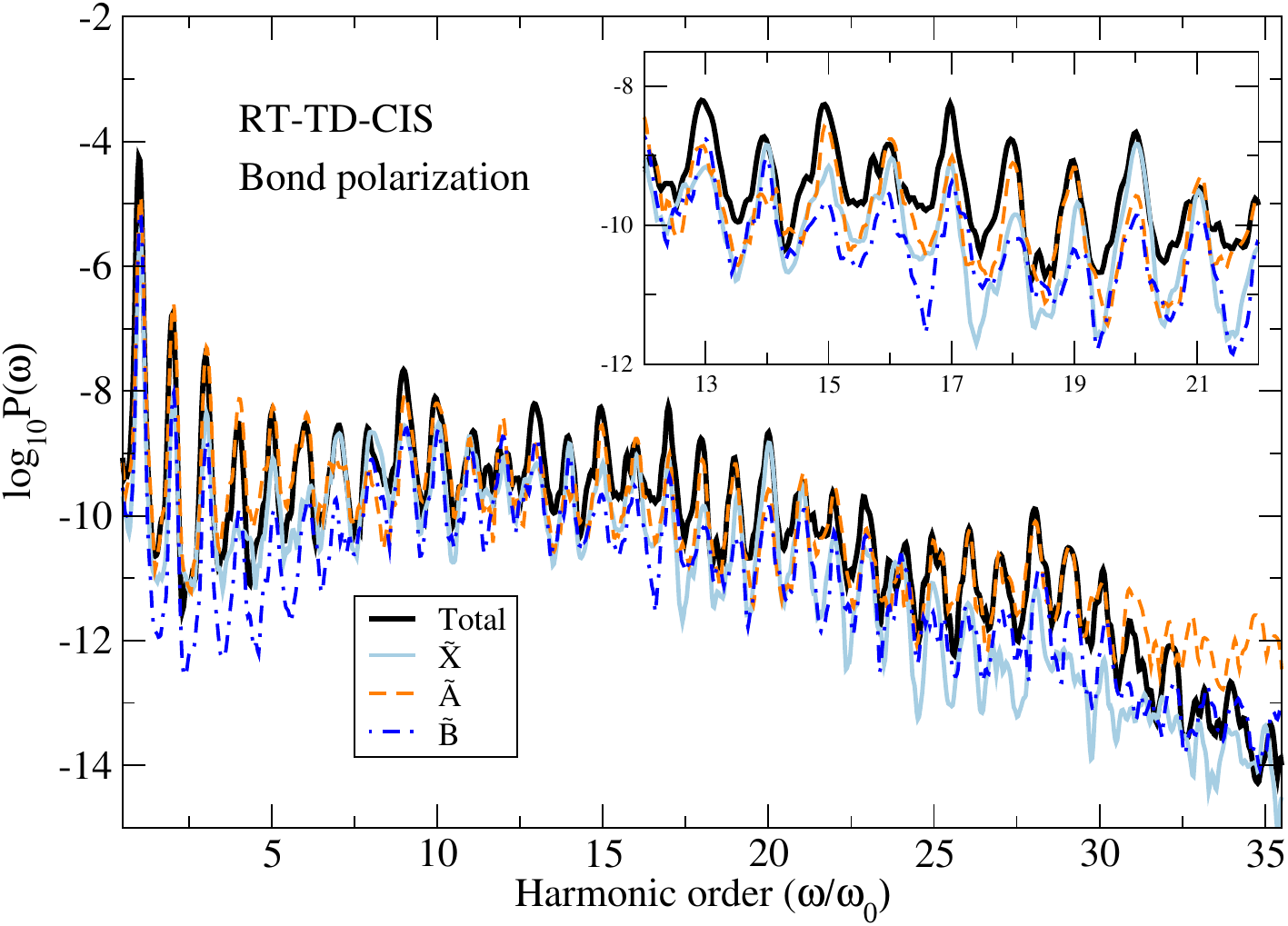}
\caption{\label{HHG_pol_L_H2O} MO decomposition of the HHG spectrum of H$_2$O, with laser-pulse polarization parallel to a O-H bond, at the RT-TD-CIS level of theory.}
\end{figure}

In Figs.~\ref{HHG_pol_Y_H2O} and \ref{HHG_pol_L_H2O} we report the MO contributions to the HHG spectra of the water molecule for the $y$ polarization and along a O-H bond, respectively (the dipole moments from each MO are collected in Figs. S14 and S15 of the SI).
The HOMO, HOMO-1, and HOMO-2 satisfy the same symmetry \cite{neu19,tzu22} with the $y$ polarization (top panel of Fig.~\ref{HHG_pol_Y_H2O}), thus giving only odd harmonics, as the full spectrum in Fig.~\ref{HHG_H2O}. 
More specifically, for $y$-polarized light, the laser pulse is polarized perpendicular to the permanent dipole of the molecule, which lies along the
$z$-axis. Because of this perpendicular orientation, odd harmonics are predominantly polarized along the $y$-axis, whereas even harmonics tend to be polarized along the $z$-axis. This is consistent with the selection rules derived from dynamical symmetries (DSs), as already mentioned \cite{neu19}. These findings are also in line with numerical simulations on similar molecules like CO and HCN, as described in Refs. \citenum{hu2017pure,chu2024orientation}. 
Thus, the symmetry of the molecular dipole dictates the distribution of odd and even harmonics, with odd harmonics aligning with the polarization of the laser pulse and even harmonics being orthogonal. This alignment behavior is shown clearly in Fig.~\ref{HHG_pol_Y_H2O} where the results match the expectations based on these selection rules.
In Fig.~\ref{HHG_pol_Y_H2O}, the channel $\tilde{\bf X}$ (HOMO) gives the most significant contribution throughout the entire energy interval considered, because of its lowest ionization potential and favorable symmetry for ionization. Indeed, the other channels $\tilde{\bf A}$ (HOMO-1) and $\tilde{\bf B}$ (HOMO-2) contribute much less to the HHG signal: they are characterized by $xy$ (HOMO-1) and $yz$ nodal planes which negatively interact with the laser pulse polarized along the $y$ axis. 
It is worth observing that $\tilde{\bf B}$ contributes even less than $\tilde{\bf A}$. This can be explained by the fact that the energy difference of $\tilde{\bf B}$ with respect to $\tilde{\bf X}$, i.e. 5.5 eV, is larger than the one of $\tilde{\bf A}$ with respect to $\tilde{\bf X}$, i.e. 2.0 eV, as reported in Tab.~\ref{tab:ip_h2o}. Indeed, the signal from $\tilde{\bf A}$ cannot be neglected at harmonics H13 and H21, and in the high-energy region. Thus, with a laser pulse polarized perpendicularly to the molecular plane, the HHG signal is strongly dominated by a single-MO dynamics, the HOMO one, even though the HOMO-1 contribution is comparable in some parts of the spectrum. 
Detecting along the 
$z$-axis, the MO decomposition reveals a dominant contribution of a specific component, denoted as $\tilde{\bf X}$. This dominant component in the spectrum is responsible for producing only even harmonics, regardless of which electronic channel is considered during the harmonic generation process. This result suggests that along the $z$-axis, the symmetry of the system and the interaction with the laser field selectively favor the emission of even harmonics, likely due to the alignment of molecular orbitals and the dipole moment along this axis. It further highlights the significant role of orbital contributions and selection rules in determining the harmonic content based on detection directions.
Conversely, the reflection symmetry is individually broken when the laser-pulse polarization is along a O-H bond direction, and consequently the partial HHG spectra for HOMO, HOMO-1, and HOMO-2 exhibit odd and even harmonics, as shown in Fig. \ref{HHG_pol_L_H2O}.

\begin{figure}[!ht]
    \centering
  
    \begin{subfigure}{.8\textwidth}
    {\includegraphics[width=\textwidth]{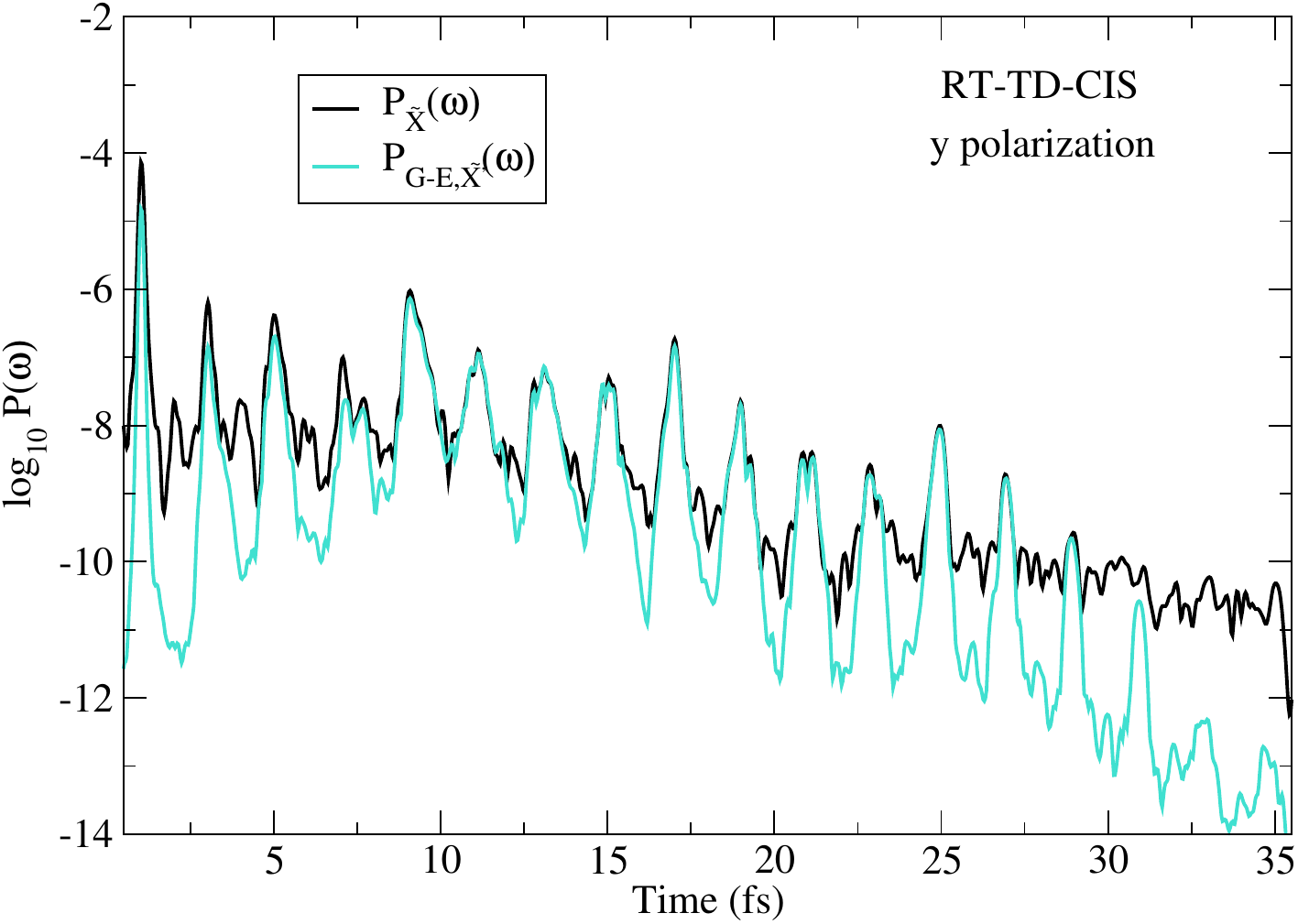}}
    \end{subfigure}
    \begin{subfigure}{.8\textwidth}
    {\includegraphics[width=\textwidth]{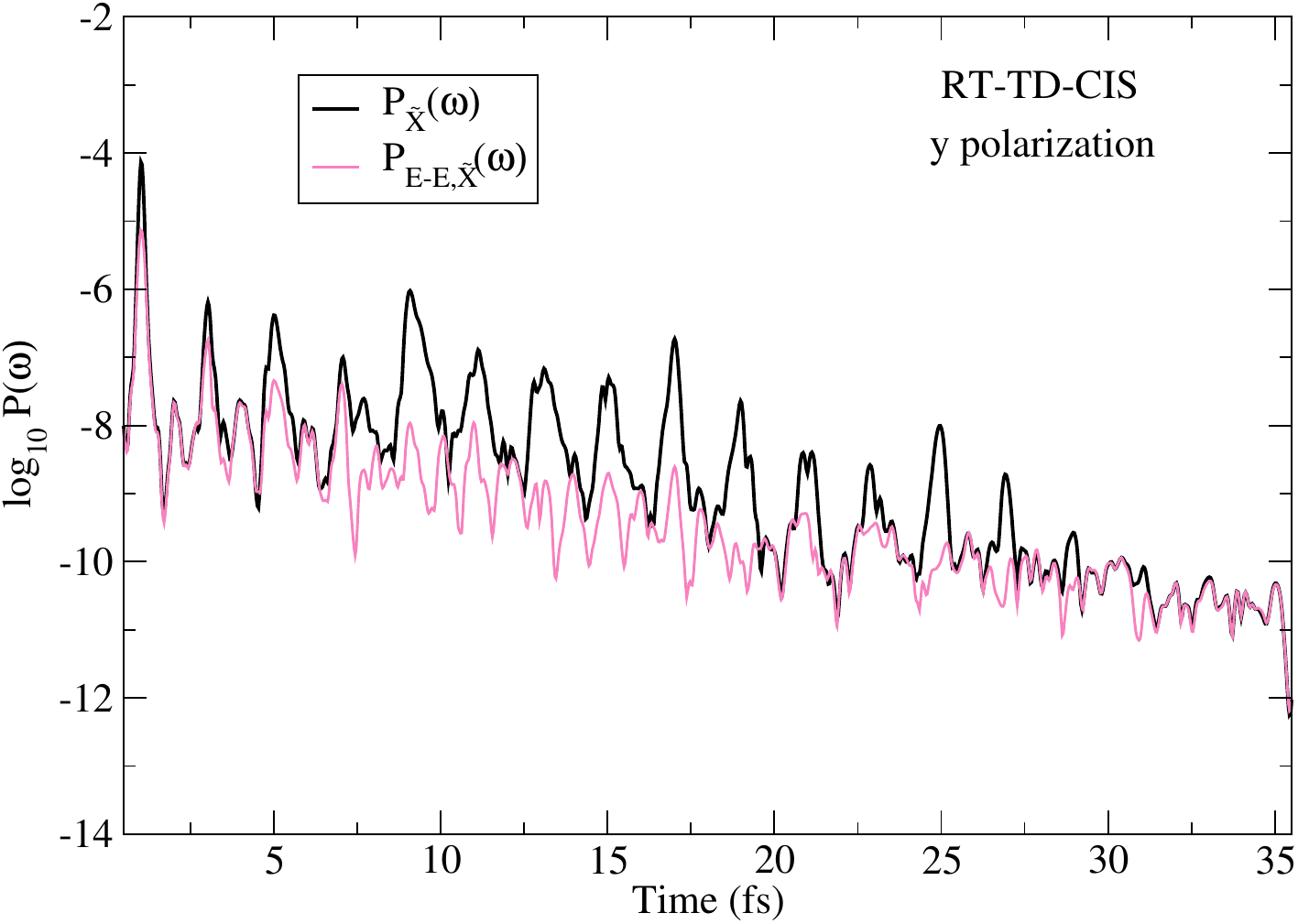}}
    \end{subfigure}
    \caption{Ground-excited (G-E, top) and excited-excited (E-E, bottom) contributions to the HHG spectrum for the $\bf X$ channel of H$_2$O,  with laser-pulse polarization perpendicular to the molecular plane, at the RT-TD-CIS level of theory.}
    \label{HHG_comp_h2o}
\end{figure}

Also for H$_2$O, the MO analysis of the HHG spectrum is enriched with the ground-excited and excited-excited contributions in Eq. (\ref{PiLambdaomega}). In Fig. \ref{HHG_comp_h2o}, we show for the channel $\tilde{\bf X}$ that the contribution $P_{i,\text{G-E}}(\omega)$ is the dominant contribution to the full spectrum $P_i(\omega)$, as already observed in CO$_2$.
The multiple-MO dynamics is much more evident in the decomposition of the HHG signal in the presence of a laser pulse linearly polarized along a O-H bond, as reported in Fig.~\ref{HHG_pol_L_H2O}.
In this case, the channel with the lowest ionization potential, $\tilde{\bf X}$, does not have a favorable symmetry, as it presents a nodal plane corresponding to the plane of the molecule. Instead, the channels $\tilde{\bf A}$  and $\tilde{\bf B}$ are characterized by a positive coupling with the laser pulse polarized along a O-H bond (see Fig.~\ref{OrbitalsCISH2O}). 
In fact, the $\tilde{\bf X}$ channel (HOMO), despite having an unfavorable symmetry for ionization along the bond direction, is easily energetically accessible (13.9 eV, Tab.~\ref{tab:ip_h2o}), making the corresponding partial signal close to the total one at low and middle-range harmonics.
The HHG signal from HOMO-1 ($\tilde{\bf A}$ channel) is the strongest almost everywhere, also at high energies.
On the other hand, the $\tilde{\bf B}$ channel (HOMO-2), which is much more difficult to ionize (19.3 eV, Tab.~\ref{tab:ip_h2o}), produces a negligible HHG signal. At variance with what is observed for the laser-pulse polarization perpendicular to the molecular plane, a clear multiple-MO signal (HOMO and HOMO-1) is obtained for water along the direction of the O-H bond. 

The same general conclusions for HOMO and HOMO-1 have been reported in Ref. \citenum{ren2023molecular}. In that work, the analysis is based on the MO symmetry and its ionization rate along a given direction: the MO with the highest ionization rate dominates the HHG signal along that direction, and the type of harmonics observed depends on the MO symmetry itself. 

  We have also computed the HHG spectra of H$_2$O along the $z$ direction, with the laser-pulse polarization along the O-H bond. Full and partial spectra are given in Fig. S16 of the SI. Odd and even harmonics are observed in this case. The channel $\tilde{\textbf{B}}$ provides a much smaller contribution than the channels $\tilde{\bf X}$ and $\tilde{\bf A}$.

%%%%%%%%%%%%%%%%%%%%%%%%%%%%%%%%%%%%%%%%%%%%%%%%%%%%%%%%%%%%%%%%
%%%%%%%%%%%%%%%%%%%%%%%%%%%%%%%%%%%%%%%%%%%%%%%%%%%%%%%%%%%%%%

%%%%%%%%%%%%%%%%%%%%%%%%%%%%%%%%%%%%%%%%%%%%%%%%%%%%%%%%%
%%%%%%%%%%%%%%%%%%%%%%%%%%%%%%%%%%%%%%%%%%%%%%%%%%%%%%%%%%%

%%%%%%%%%%%%%%%%%%%%%%%%%%%%%%%%%%%%%%%%%%%%%%%%%%%%%%%%%%%
%%%%%%%%%%%%%%%%%%%%%%%%%%%%%%%%%%%%%%%%%%%%%%%%%%%%%%%%%ù

\section{Conclusions} 
\label{con}

We propose a MO decomposition within the framework of RT-TD-CIS to extract the contributions of each MO to the HHG spectrum. The interpretation of the MO role in HHG is based on the analysis of the orbital's symmetry and energy. We studied the CO$_{2}$ and H$_{2}$O molecules using this approach.

The HHG spectrum of CO$_2$ is well-established to exhibit a dynamic minimum, arising from destructive interferences between ionization channels, particularly between molecular orbitals such as the HOMO and HOMO-2. This minimum, typically observed around the 23rd harmonics, is sensitive to the laser parameters such as intensity and polarization, reflecting complex multi-channel dynamics. Our RT-TD-CIS methodology successfully reproduces this minimum at H23 (see Fig. 3), showing that channel $\textbf{B}$ (HOMO-2) is populated similarly to channel $\textbf{X}$ (HOMO) due to its favorable symmetry. A similar trend is observed for channel $\textbf{C}$ (HOMO-3), especially at higher harmonics. The dynamic minimum is accurately captured, with clear evidence that its appearance results from a destructive interference between channels $\textbf{X}$ and $\textbf{B}$ \cite{Smirnova2009}. When the polarization is perpendicular, channel $\textbf{A}$ plays a significant role in modulating the HHG signal. Using the RT-TD-CIS-LC-$\omega$PBE method, the overall picture remains consistent, despite differences in orbital level ordering. By employing two distinct methods for the electronic structure and the strong-field dynamics, we demonstrate the robustness of our approach and confirm that the underlying physics of HHG is preserved, regardless of variations in level ordering.

For aligned H$_{2}$O, the HHG spectrum with a pulse polarization along the O-H bond is much weaker than that with a $y$ polarization. In the first case, also even harmonics appear, due to symmetry breaking, and the HHG spectrum is also much weaker. The channel $\tilde{\textbf{X}}$ (HOMO) provides the largest contribution with the laser-pulse polarization along the $y$ axis. Instead, a clear multiple-orbital HHG signal is observed along the O-H bond, involving the channels $\tilde{\textbf{X}}$ and ${\textbf{A}}$ (HOMO-1).

Our methodology enables a real-time molecular orbital decomposition within wave-function methods under strong-field conditions, offering significant advantages over traditional techniques, such as the ability to efficiently explore large biomolecules through Gaussian basis sets.

\begin{acknowledgement}
Financial support from ICSC – Centro Nazionale di Ricerca in High Performance Computing, Big Data and Quantum Computing, funded by European Union – NextGenerationEU is gratefully acknowledged.
This work has been supported by the project CHANGE funded by the PRIN 2022 - Progetti di Rilevante Interesse Nazionale (grant 20224KAC28).
\end{acknowledgement}

\begin{suppinfo}
MO contributions to time-dependent dipole moments for CO$_2$ and H$_2$O. Velocity and acceleration HHG spectra for CO$_2$ and laser-pulse polarization along $z$. Interference magnitudes for CO$_2$ with laser-pulse polarization along $z$. Time-dependent dipole moments and HHG spectra together with MO contribution for CO$_2$ molecule at the RT-TD-CIS-LC-$\omega$PBE level of theory. Comparison between total HHG spectra of CO$_2$ at the RT-TD-CIS and  RT-TD-CIS-LC-$\omega$PBE level of theory. MO decomposition of the time-dependent dipole moments for H$_2$O at the RT-TD-CIS level of theory.
\end{suppinfo}

\section*{Data Availability Statement}

Data available on request from the authors.

%\bibliography{bib_rev}% Produces the bibliography via BibTeX.

\providecommand{\latin}[1]{#1}
\makeatletter
\providecommand{\doi}
  {\begingroup\let\do\@makeother\dospecials
  \catcode`\{=1 \catcode`\}=2 \doi@aux}
\providecommand{\doi@aux}[1]{\endgroup\texttt{#1}}
\makeatother
\providecommand*\mcitethebibliography{\thebibliography}
\csname @ifundefined\endcsname{endmcitethebibliography}
  {\let\endmcitethebibliography\endthebibliography}{}

\end{document}